\documentclass{article} 

\usepackage[english]{babel} 
\usepackage{amssymb}
\usepackage{amsmath}
\usepackage{txfonts}
\usepackage{multirow}

\usepackage{mathdots}
\usepackage[classicReIm]{kpfonts}
\usepackage[pdftex]{graphicx} 

\begin{document}

\noindent \includegraphics*[width=2.42in, keepaspectratio=true]{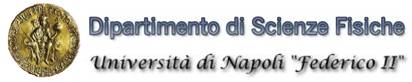}\includegraphics*[width=2.25in,  keepaspectratio=true]{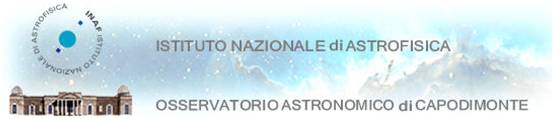}\includegraphics*[width=1.20in,  keepaspectratio=true]{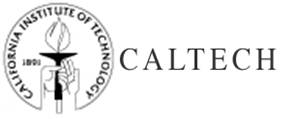}

\vskip 5cm
\noindent
\begin{center}

 \textit{{\huge DAMEWARE}}
\vskip 1cm
 \textit{{\huge Data Mining \& Exploration }}

 \textit{{\huge Web Application Resource}}
\end{center}
\noindent

\vskip 4cm
\noindent

\noindent
\begin{center}
\begin{tabular}{|p{.6in}|p{3.5in}|} \hline
Issue:  & 1.5 \\ \hline
Date:  & March 1, 2016 \\ \hline
Authors: & M. Brescia, S. Cavuoti, F. Esposito, M. Fiore, M. Garofalo, M. Guglielmo, G. Longo, F. Manna, A. Nocella, C. Vellucci \\ \hline
\end{tabular}
\end{center}
\vskip 2cm

\noindent \includegraphics*[width=\textwidth,  keepaspectratio=true]{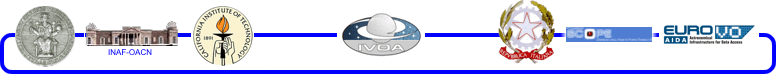}

\noindent \textbf{\textit{}}
\begin{center}

\noindent \textbf{\textit{DAME Program}}

\noindent \textbf{\textit{``we make science discovery happen''}}
\end{center}

\newpage

\tableofcontents


\listoftables




\listoffigures



\newpage
\section{ Abstract}

\noindent Astronomy is undergoing through a methodological revolution triggered by an unprecedented wealth of complex and accurate data. DAMEWARE (DAta Mining \& Exploration Web Application and REsource) is a general purpose, Web-based, Virtual Observatory compliant, distributed data mining framework specialized in massive data sets exploration with machine learning methods. We present the DAMEWARE (DAta Mining \& Exploration Web Application REsource) which allows the scientific community to perform data mining and exploratory experiments on massive data sets, by using a simple web browser. DAMEWARE offers several tools which can be seen as working environments where to choose data analysis functionalities such as clustering, classification, regression, feature extraction etc., together with models and algorithms.

\section{ Introduction}

\noindent The present document is part of the DAMEWARE \cite{4,10} Web Application Suite user-side documentation package. The final release arises from the very primer version of the web application ($\alphaup$ release) which has been made available to public domain since July 2010. Currently this release has been more updated, by fixing residual bugs and by adding more functionalities and models.

\noindent The final developing team has spent much efforts to fix bugs, satisfy testing user requirements, suggestions and to improve the application features, by integrating several other data mining models, always coming from machine learning theory, which have been scientifically validated by applying them offline in several practical astrophysical cases (photometric redshifts, quasar candidate selection, globular cluster search, transient discovery etc). Some examples of such scientific validation can be found in:
\cite{2,3,5,6,8,9,10,11,12,14,16,22,24}. All cases dealing with time domain data rich astronomy. In this scenario, the $\alphaup$ release has covered the role of an advanced prototype, useful to evaluate, tune and improve main features of the web application, basically in terms of:

\noindent

\begin{enumerate}
\item  User friendliness: by taking care of the impact on new users, not necessarily expert in data mining or skilled in machine learning methodologies, by paying particular attention to the easiness of navigation through GUI options and to the learning speed in terms of experiment selection, preprocessing, setup and execution;

\item  Data I/O handling: easiness to upload/download data files, to edit and configure datasets from original data files and/or archives;

\item  Workspace handling: the capacity to create different work spaces, depending on the experiment type and data mining model choice;
\end{enumerate}

\noindent

\noindent Of course in the new release it was impossible to match all important and valid suggestions came from the $\alphaup$ release testers. In principle not for bad will of developers, but mostly because in some cases, the requests would needed drastic re-engineering of some infrastructure components or simply because they went against our design requirements, issued at the very beginning of the project. Of course, this not implies necessarily that in next releases of the application these requests will not be taken into account.

\noindent Anyway, we tried to satisfy as much as possible main requests concerning the improvement of ease to use. Also in terms of examples and guided tours in using the available models. Don't forget that neophyte users should spend a certain amount of time to read this and other manuals to learn their capabilities and usability topics before to move inside the application. This is particularly true in order to understand how to identify the right association of functionality domain and the data mining model to be applied to your own science case. But we recall that this is fully reachable by gaining experience with time and through several trial-and-error sessions. DAMEWARE offers also the possibility to extend the original library of available tools by allowing end users to plug-in\footnote{ $  $http://dame.dsf.unina.it/dameware.html\#plugin }, expose to the community, and execute their own code in a simple way  by downloading a plugin wizard, configuring his program I/O interface and then uploading the plugin which is automatically created. All this without any restriction about the native programming language.

\section{ Purpose}

\noindent This manual is mainly dedicated to drive users through the GUI options and features. In other words to show how to navigate and to interact with the application interface in order to create working spaces, experiments, to upload/download and edit data files. We will stop our discussion here at level of configuration of the models, for which specific manuals are available. This in order to separate the use of the GUI from the theoretical implications related to the setup and use of available data mining models.

\noindent \textbf{The access gateway, its complete documentation package and other resources is at the following address:}

\noindent http://dame.dsf.unina.it/dameware.html

\noindent

\noindent Last pages of this document host the table with Abbreviations \& Acronyms, Reference lists and the acknowledgments.

\section{ GUI Overview}

\noindent Main philosophy behind the interaction between user and the DMS (Data Mining Suite) is the following.

\noindent The DMS is a web application, accessible through a simple web browser. It is structurally organized under the form of working sessions (hereinafter named workspaces) that the user can create, modify and erase. You can imagine the entire DMS as a container of services, hierarchically structured as in Fig. 1. The user can create as many workspaces as desired and populate them with uploaded data files and with experiments (created and configured by using the Suite). Each workspace is enveloping a list of data files and experiments, the latter defined by the combination between a functionality domain and a series (one at least) of data mining models. From these considerations, it is obvious that a workspace makes sense if at least one data file is uploaded into.
\textbf{So far, the first two actions, after logged in, are, respectively, to create a new workspace (by assigning it a name) and to populate it by uploading at least one data file, to be used as input for future experiments. The data file types allowed by the DMS are reported in the next sections.}

\noindent In principle there should be many experiments belonging to a single workspace, made by fixing the functional domain and by slightly different variants of a model setup and configuration or by varying the associated models.

\noindent

\begin{figure}  \centering\includegraphics*[width=3.09in,  keepaspectratio=true]{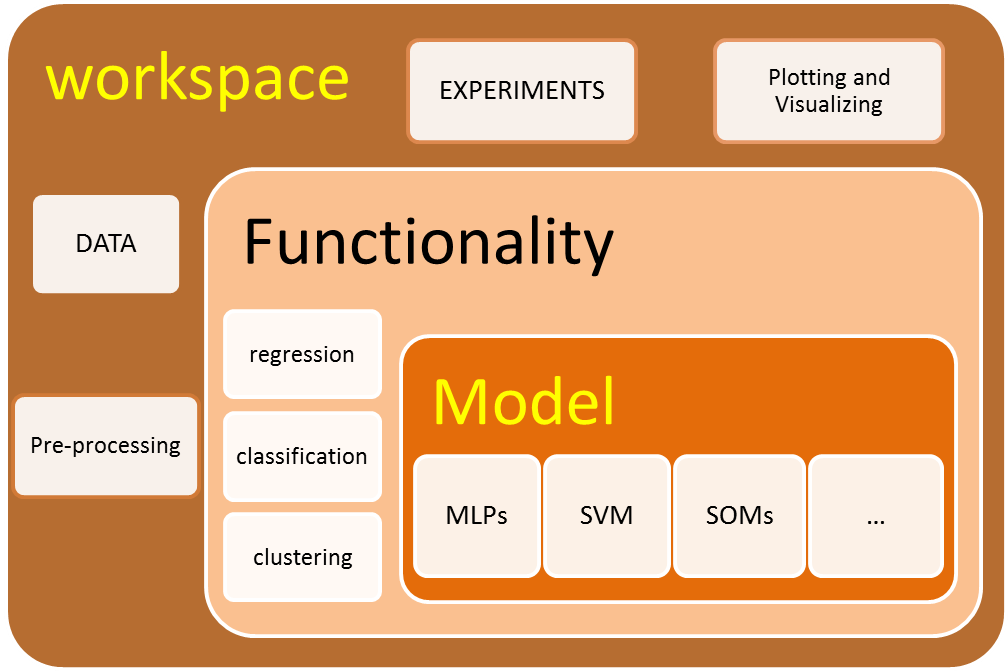}

\caption{Suite functional hierarchy}\end{figure}

\noindent

\noindent By this way, as usual in data mining, the knowledge discovery process should basically consist of several experiments belonging to a specified functionality domain, in order to find the model, parameter configuration and dataset (parameter space) choices that give the best results (in terms of performance and reliability). The following sections describes in detail the practical use of the DMS from the end user point of view. Moreover, the DMS has been designed to build and execute a typical complete scientific pipeline (hereinafter named workflow) making use of machine learning models. This specification is crucial to understand the right way to build and configure data mining experiment with DMS.

\noindent In fact, machine learning algorithms (hereinafter named models) need always a pre-run stage, usually defined as training (or learning phase) and are basically divided into two categories: supervised and unsupervised models, depending, respectively, if they make use of a BoK (Base of Knowledge), i.e. couples input-target for each datum, to perform training or not (for more details about the concept of training data, see section 3.5 below).

\noindent So far, any scientific workflow must take into account the training phase inside its operation sequence.

\noindent Apart from the training step, a complete scientific workflow always includes a well-defined sequence of steps, including pre-processing (or equivalently preparation of data), training, validation, run, and in some cases post-processing.

\noindent

\noindent The DMS permits to perform a complete workflow, having the following features:

\noindent

\begin{enumerate}
\item  A workspace to envelope all input/output resources of the workflow;

\item  A dataset editor, provided with a series of pre-processing functionalities to edit and manipulate the raw data uploaded by the user in the active workspace (see section 4.4 for details);

\item  The possibility to copy output files of an experiment in the workspace to be arranged as input dataset for subsequent execution (the output of training phase should become the input for the validate/run phase of the same experiment);

\item  An experiment setup toolset, to select functionality domain and machine learning models to be configured and executed;

\item  Functions to visualize graphics and text results from experiment output;

\item  A plugin-based toolkit to extend DMS functionalities and models with user own applications;
\end{enumerate}

\noindent

\begin{figure}  \centering\includegraphics*[width=\textwidth]{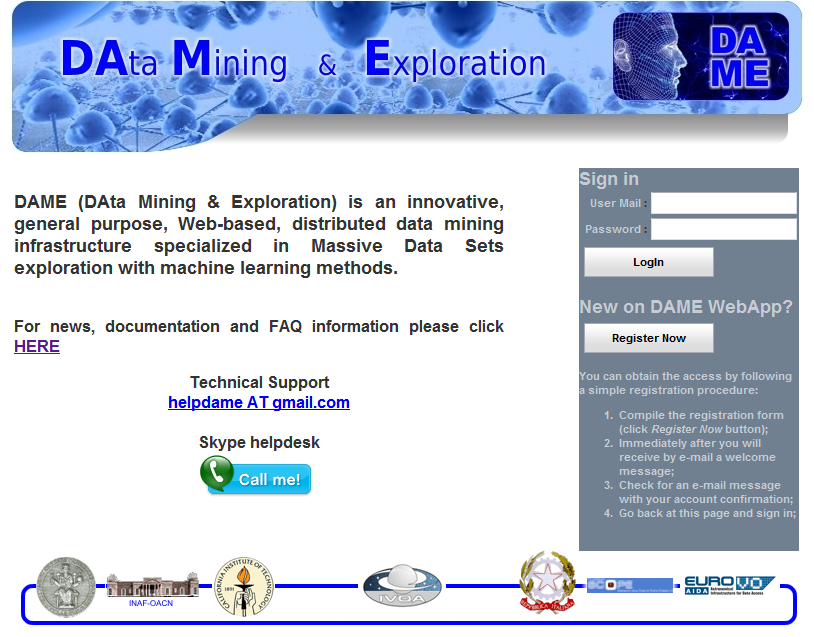}

\caption{The user registration/login form to access at the web application}
\end{figure}
\noindent

\noindent New users must be registered by following a very simple procedure requiring to select ``Register Now'' button on that page.

\noindent The registration form requires the following information to be filled in by the user (all fields are required):

\noindent

\begin{enumerate}
\item  Name of the user;

\item  Family name of the user;

\item  User e-mail: the user e-mail (it will become his access login). It is important to define a real address, because it will be also used by the DMS for communications, feedbacks and activation instructions;

\item  Country: country of the user;

\item  Affiliation: the institute/academy/society of the user;

\item  Password: a safe password (at least 6 chars), without spaces and special chars;
\end{enumerate}

\noindent
\begin{figure}
\includegraphics*[width=\textwidth]{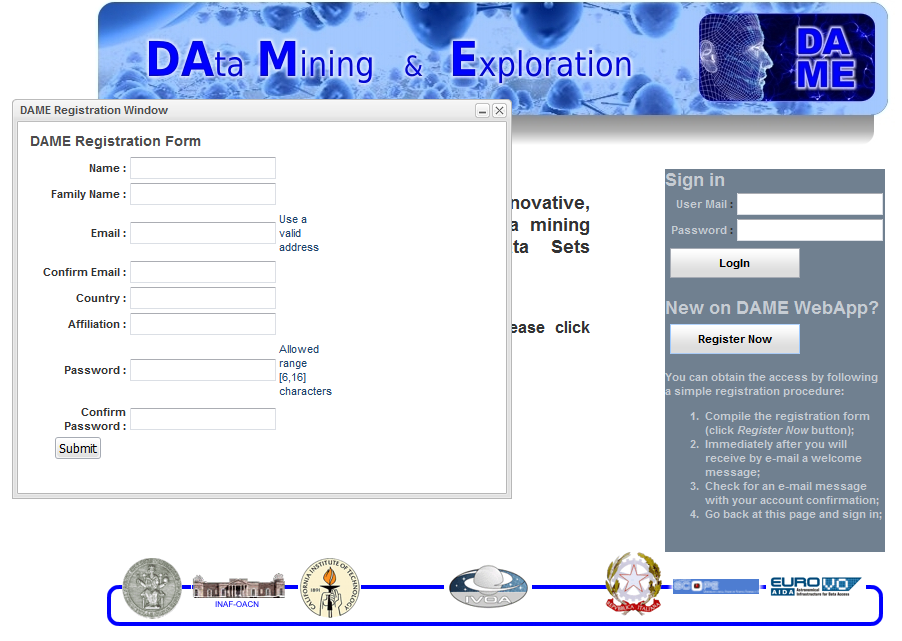}

\caption{The user registration form}
\end{figure}
\noindent

\noindent After submission, an e-mail will be immediately sent at the defined address (Fig. 4), confirming the correct coming up of the activation procedure.

\noindent

\noindent
\begin{figure}  \centering\includegraphics*[width=\textwidth,  keepaspectratio=true]{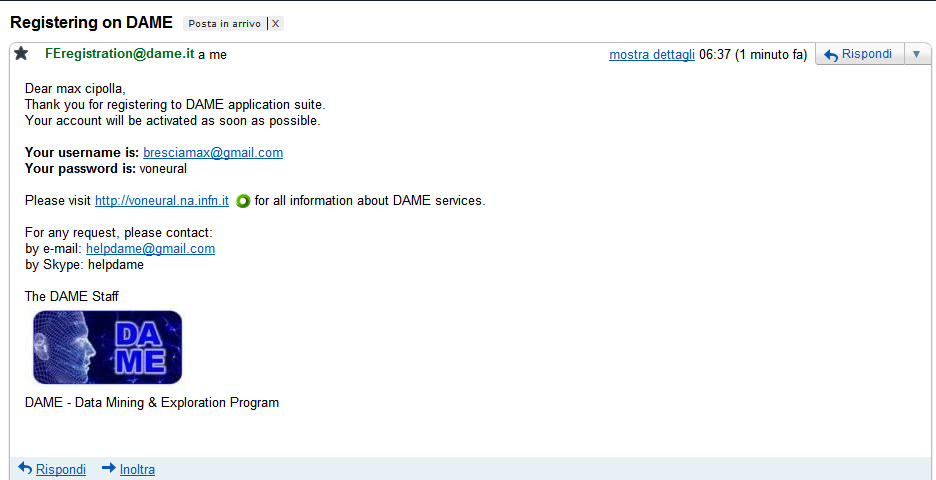}
\caption{An example of e-mail received by the user after submission of registration info}
\end{figure}
\noindent

\noindent

\noindent After that the user must wait for a second e-mail which will be the final confirmation about the activation of the account. This is required in order to provide an higher security level.

\noindent Once the user has received the activation confirmation, he can access the webapp by inserting e-mail address and password.

\noindent The webapp will appear as shown in Fig. 5

\noindent

\begin{figure}  \centering\includegraphics*[width=\textwidth]{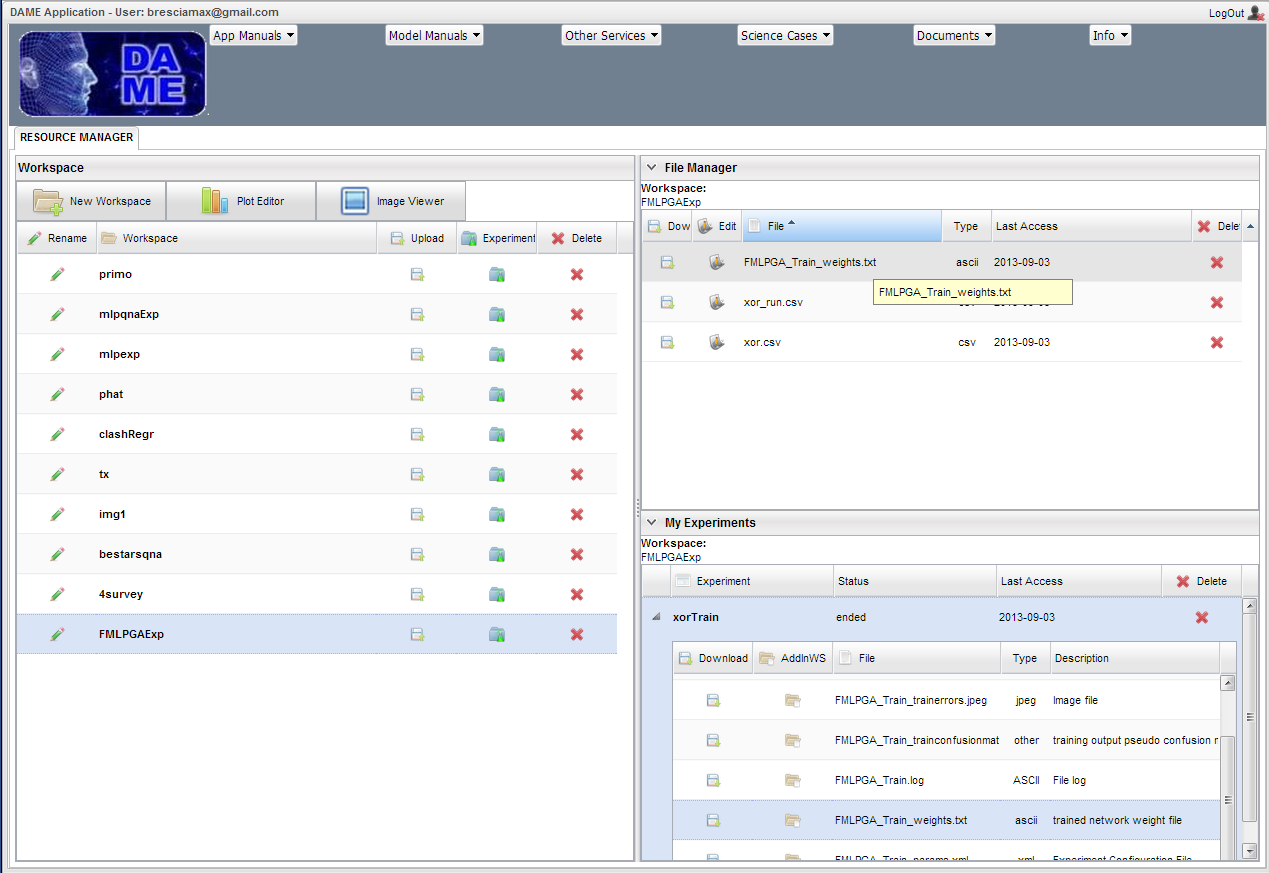}

\caption{The Web Application starting main page (Resource Manager)}
\end{figure}

\subsection{ The command icons}

\noindent The interaction between user and GUI is based on the selection of icons, which correspond to basic features available to perform actions. Here their description, related to the red circles in Fig. 6 is reported:

\noindent

\begin{enumerate}
\item  \textbf{The header menu options:} When one of the available menus is selected, a drop  submenu appears with some options;

\item  \textbf{Logout button}: If pressed the GUI (and related working session) is closed;

\item  \textbf{Operation tabs:} The GUI is organized like a multi-tab browser. Different tabs are automatically open when user wants to edit data file to create/manipulate datasets, to upload files or to configure and launch experiments. All tabs can be closed by user, except the main one (Resource Manager);

\item  \textbf{Creation of new workspaces}: When selected and named, the new workspace appears in the Workspace List Area (Workspace sub window);

\item  \textbf{Workspace List Area}: portion of the main Resource Manager tab dedicated to host all user defined workspaces;

\item  \textbf{Upload command:} When selected, the user is able to select a new file to be uploaded into the Workspace Data Area (Files Manager sub window). The file can be uploaded from external URI or from local (user) HD;

\item  \textbf{Creation of new experiment}: When selected, the user is able to create a new experiment (a specific new tab is open to configure and launch the experiment);

\item  \textbf{Rename workspace command:} When selected the user can rename the workspace;

\item  \textbf{Delete Workspace command}: When selected, the user can delete the related workspace (only if no experiments are present inside, otherwise the system alerts to empty the workspace before to erase it);

\item  \textbf{File Manager Area}: the portion of Resource Manager tab dedicated to list the data files belonging to various workspaces. All files present in this area are considered \underbar{as input files} for any kind of experiment;

\item  \textbf{Download command}: When selected the user can download locally (on his HD) the selected file;

\item  \textbf{Dataset Editor command:} When selected a new tab is open, where the user can create/editdataset files by using all available dataset manipulation features;

\item  \textbf{Delete file command}: When selected the user can delete the selected file from current workspace;

\item  \textbf{Experiment List Area}: The portion of Resource Manager tab dedicated to the list of experiments and related output files present in the selected workspace;

\item  \textbf{Experiment verbose list command}: When selected the user can open the experiment file list (for experiment in ended state) in a verbose mode, showing all related files created and stored;

\item  \textbf{Delete Experiment command}: by clicking on it, the entire experiment (all listed files) is erased;

\item  \textbf{Download experiment file command:} When selected the user can download locally (on his HD) the related experiment output file;

\item  \textbf{AddinWS command}: When selected, the related file is automatically copied from the Experiment List Area to the currently active workspace File Manager Area. This feature is useful to re-use an output file of a previous experiment as input file of a new experiment (in the figure, look at the file weights.txt, that after this command is also listed in the File Manager). A file present in both areas, can be used as input either as output in the experiments;

\item  \textbf{Plot Editor: }When pressed open in the resource manager four tabs: histogram, scatter 2d, scatter 3d and line plot; each tab is dedicated to a specific type of plot;

\item  \textbf{Image Viewer: }When pressed open a new tab in the resource manager dedicated to the visualization of an image. The image file is intended already loaded in the File Manager.
\end{enumerate}

\noindent

\begin{figure}\includegraphics*[width=\textwidth]{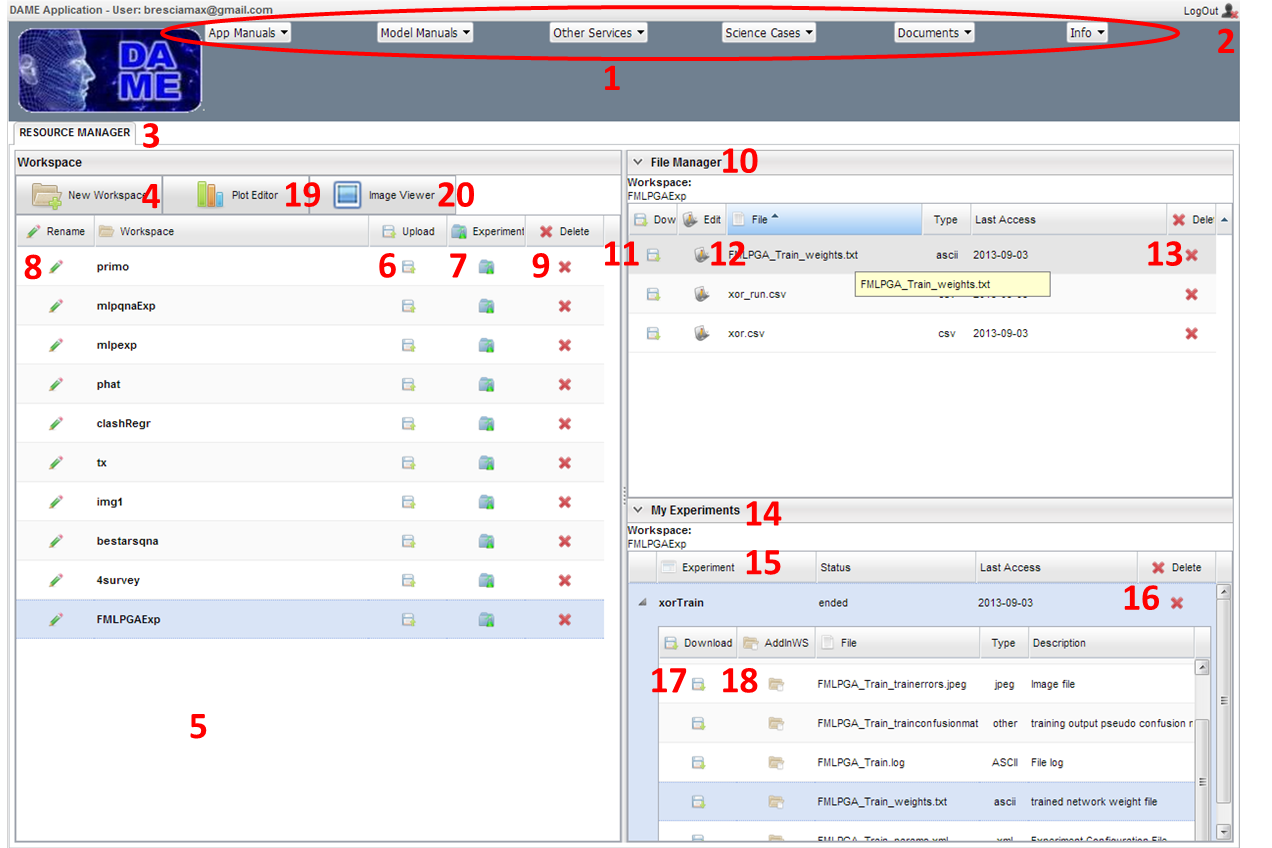}

\caption{The Web Application main areas and commands}
\end{figure}

\subsection{ Workspace Management}

\noindent A workspace is namely a working session, in which the user can enclose resources related to scientific data mining experiments. Resources can be data files, uploaded in the workspace by the user, files resulting from some manipulations of these data files, i.e. dataset files, containing subsets of data files, selected by the user as input files for his experiments, eventually normalized or re-organized in some way (see section4.4 for details). Resources can also be output files, i.e. obtained as results of one or more experiments configured and executed in the current ``active'' workspace (see section 4.6for details).

\noindent The user can create a new or select an existing workspace, by specifying its name. After opening the workspace, this automatically becomes the ``active'' workspace. This means that any further action, manipulating files, configuring and executing experiments, upload/download files, will result in the active workspace, Fig. 7. In this figure it is also shown the right sequence of main actions in order to operate an experiment (workflow) in the correct way.

\noindent

\begin{figure}  \centering\includegraphics*[width=4.36in]{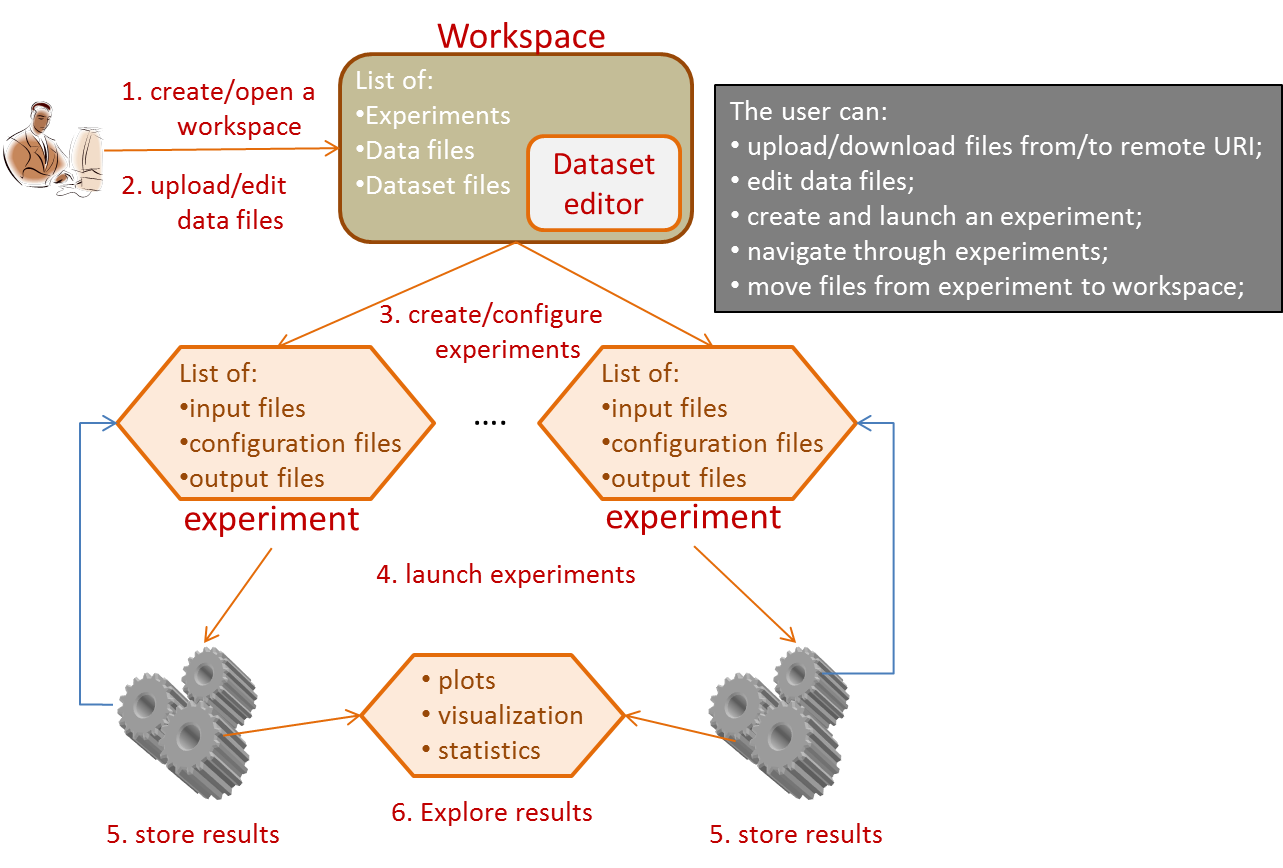}

\caption{The right sequence to configure and execute an experiment workflow}
\end{figure}
\noindent

\noindent So far, the basic role of a workspace is to make easier to the user the organization of experiments and related input/output files. For example the user could envelope in a same workspace all experiments related to a particular functionality domain, although using different models.

\noindent It is always possible to move (copy) files from experiment to workspace list, in order to re-use a same dataset file for multiple experiment sessions, i.e. to perform a workflow.

\noindent After access, the user must select the ``active'' workspace. If no workspaces are present, the user must create a new one, otherwise the user must select one of the listed workspace. The user can always create a new workspace by pressing the button as in Fig. 8.

\noindent

\begin{figure}  \centering\includegraphics*[width=1.50in]{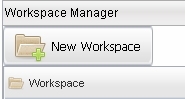}

\caption{the button ``New Workspace'' at left corner of workspace manager window}
\end{figure}
\noindent

\noindent As consequence the user must assign a name to the new workspace, by filling in the form field as in Fig. 9.

\noindent

\begin{figure}  \centering\includegraphics*[width=2.01in]{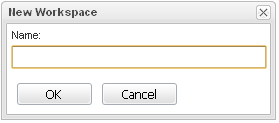}

\caption{the form field that appears after pressing the ``New Workspace'' button}
\end{figure}
\noindent

\noindent After creation, the active workspace can be populated by data and experiments, Fig. 10.

\noindent

\begin{figure}  \centering\includegraphics*[width=\textwidth]{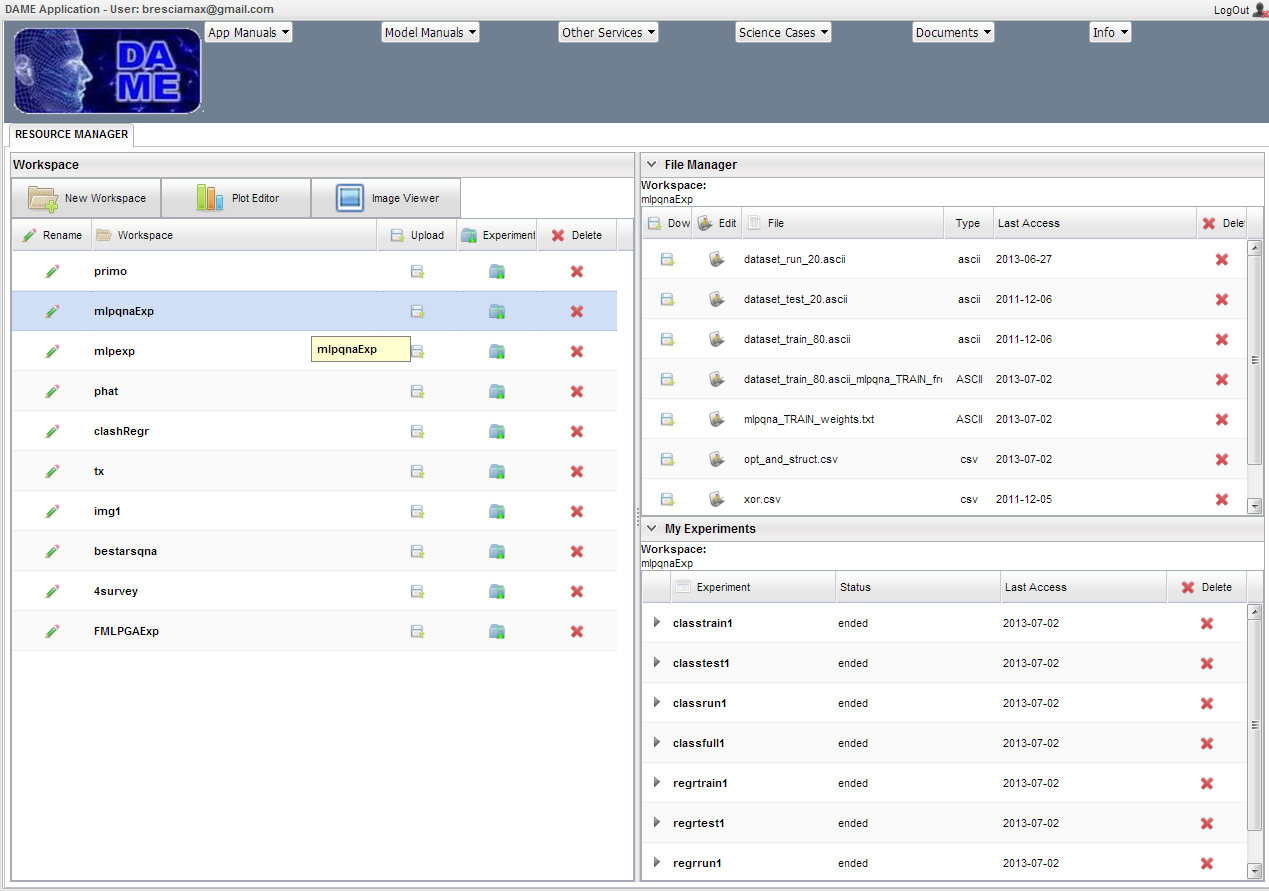}

\caption{the active workspace created in the Workspace List Area}
\end{figure}

\subsection{ Header Area}

\noindent At the top segment of the DMS GUI there is the so-called Header Area. Apart from the DAME logo, it includes a persistent menu of options directly related to information and documentation (this document also) available online and/or addressable through specific DAME program website pages.

\noindent

\begin{figure}  \centering\includegraphics*[width=\textwidth]{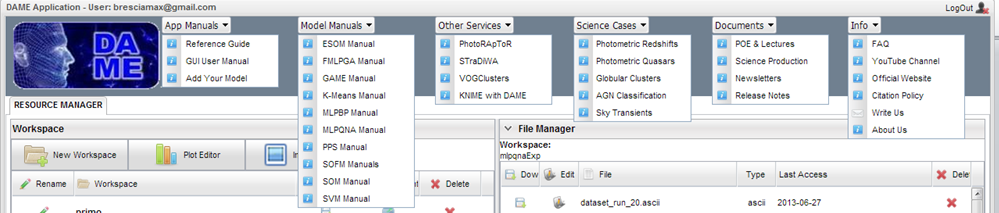}

\caption{The GUI Header Area with all submenus open}
\end{figure}
\noindent

\noindent The options are described in the following table (Tab. 1).

\noindent
\begin{table}\scriptsize
\begin{tabular}{|c|c|c|} \hline
\textbf{OPTIONS} & \textbf{HEADER} & \textbf{DESCRIPTION} \\ \hline
Reference Guide & \multirow{3}{*}{\textbf{Application Manuals}} & http://dame.dsf.unina.it/dameware.html\#appman \\
GUI User Manual &  &  \\
Extend DAME &  & http://dame.dsf.unina.it/dameware.html\#plugin \\ \hline
 &  \multirow{13}{*}{\textbf{Model Manuals}} & \textbf{Specific data mining model user manuals for experiments} \\
ESOM  Manual & & http://dame.dsf.unina.it/dameware.html\#manuals  \\
FMLPGA Manual &  & http://dame.dsf.unina.it/dameware.html\#manuals  \\
GAME & & http://dame.dsf.unina.it/dameware.html\#manuals \\
Random Forest Manual &  & http://dame.dsf.unina.it/dameware.html\#manuals  \\
K-Means Manual &  & http://dame.dsf.unina.it/dameware.html\#manuals  \\
MLPBP Manual &  & http://dame.dsf.unina.it/dameware.html\#manuals  \\
MLPQNA/LEMON Manual &  & http://dame.dsf.unina.it/dameware.html\#manuals  \\
PPS Manual &  & http://dame.dsf.unina.it/dameware.html\#manuals  \\
SOFM Manual &  & http://dame.dsf.unina.it/dameware.html\#manuals  \\
SOM Manual &  & http://dame.dsf.unina.it/dameware.html\#manuals  \\
SVM Manual &  & http://dame.dsf.unina.it/dameware.html\#manuals  \\ \hline
PhotoRApToR &  \multirow{4}{*}{\textbf{Other Services}} & http://dame.dsf.unina.it/dame\_photoz.html\#photoraptor \\
STraDIWA &  & http://dame.dsf.unina.it/dame\_td.html \\
VOGCLUSTERS App &  & http://dame.dsf.unina.it/vogclusters.html \\
KNIME with DAME &  & http://dame.dsf.unina.it/dame\_kappa.html \\ \hline
Photometric Redshifts & \multirow{5}{*}{\textbf{Science Cases}} & http://dame.dsf.unina.it/dame\_photoz.html \\
Photometric Quasars &  & http://dame.dsf.unina.it/dame\_qso.html \\
Globular Clusters &  & http://dame.dsf.unina.it/dame\_gcs.html \\
AGN Classification &  & http://dame.dsf.unina.it/dame\_agn.html \\
Sky Transients &  & http://dame.dsf.unina.it/dame\_td.html \\ \hline
POE \& Lectures & \multirow{4}{*}{\textbf{Documents}} & http://dame.dsf.unina.it/documents.html \\
Science Production & & http://dame.dsf.unina.it/science\_papers.html \\
Newsletter &  & http://dame.dsf.unina.it/newsletters.html \\
Release Notes &  & http://dame.dsf.unina.it/dameware.html\#notes \\ \hline
FAQ & \multirow{6}{*}{\textbf{Info}} & http://dame.dsf.unina.it/dameware.html\#faq \\
YouTube Channel & & http://www.youtube.com/user/DAMEmedia \\
Official website & & http://dame.dsf.unina.it \\
Citation Policy & & http://dame.dsf.unina.it/\#policy \\
Write Us & & helpdame@gmail.com \\
About Us &  & http://dame.dsf.unina.it/project\_members.html \\ \hline
\end{tabular}

\caption{Header Area Menu Options}
\end{table}

\subsection{ Data Management}

\noindent The Data are the heart of the web application (data mining \& exploration). All its features, directly or not, are involved within the data manipulation. So far, a special care has been devoted to features giving the opportunity to upload, download, edit, transform, submit, create data.

\noindent In the GUI input data (i.e. candidates to be inputs for scientific experiments) are basically belonging to a workspace (previously created by the user). All these data are listed in the ``Files Manager'' sub window. These data can be in one of the supported formats, i.e. data formats recognized by the web application as correct types that can be submitted to machine learning models to perform experiments. They are:

\noindent

\begin{enumerate}
\item  \textbf{FITS (tabular and image .fits files);}

\item \textbf{ ASCII (.txt or .dat ordinary files);}

\item \textbf{ VOTable (VO compliant XML document files);}

\item \textbf{ CSV (Comma Separated Values .csv files);}

\item \textbf{ JPEG, GIF and PNG images.}
\end{enumerate}

\noindent

\noindent The user has to pay attention to use input data in one of these supported formats in order to launch experiments in a right way.

\noindent

\noindent Other data types are permitted but not as input to experiments. For example, log, jpeg or ``not supported'' text files are generated as output of experiments, but only supported types can be eventually re-used as input data for experiments.

\noindent There is an exception to this rule for file format with extension \textbf{.ARFF (Attribute Relation File Format)}. These files can be uploaded and also edited by dataset editor, by using the type ``CSV''. But their extension .ARFF is considered ``unsupported'' by the system, so you can use any of the dataset editor options to change the extension (automatically assigned as CSV). Then you can use such files as input for experiments.

\noindent These output file are generally listed in the ``Experiment Manager'' sub window, that can be verbosely open by the user by selecting any experiment (when it is under ``ended'' state).

\noindent Other data files are created by dataset creation features, a list of operations that can be performed by the user, starting from an original data file uploaded in a workspace. These data files are automatically generated with a special name as output of any of the manipulation dataset operations available.

\noindent

\noindent \textbf{Besides these general rules, there are some important prescriptions to take care during the preparation of data to be submitted and the setup of any Machine Learning model:}

\noindent

\begin{enumerate}
\item  \textbf{Input features to any machine learning model must be scalars, not arrays of values or chars. In case, you could try to find numerical representation of any not scalar or alphanumerical quantities;}
\item \textbf{ The input layer of a generic hierarchical neural network must be populated according to the number of physical input features of your table entries. There must be a perfect correspondence between number of input nodes and input features (columns of your table);}
\item \textbf{ All objects (rows) of an input table must have exactly the same number of columns. No rows with variable number of columns are allowed;}
\end{enumerate}

\noindent \textbf{}

\begin{enumerate}
\item \textbf{ Hidden layers of any multi-layer feed-forward model (i.e. layers between input and output ones) must contain a decreasing number of nodes, usually by following an empirical law: given N input nodes, the first hidden layer should have 2N+1 nodes at least; the optional second hidden layer N-1 and so on...;}
\end{enumerate}

\noindent \textbf{}

\begin{enumerate}
\item \textbf{ For most of the available neural networks models (with exceptions of Random Forest and SVM), the class target column should be encoded with a binary representation of the class label. For example, if you have 3 different classes, you must create three different columns of targets, by encoding the 3 classes as, respectively: 001, 010, 100.}
\end{enumerate}

\noindent

\noindent Confused? Well, don't panic please. Let's read carefully next sections.

\paragraph{ Upload user data}

\noindent As mentioned before, after the creation of at least one workspace, the user would like to populate the workspace with data to be submitted as input for experiments. Remember that in this section we are dealing with supported data formats only!

\noindent

\begin{figure}  \centering\includegraphics*[width=\textwidth]{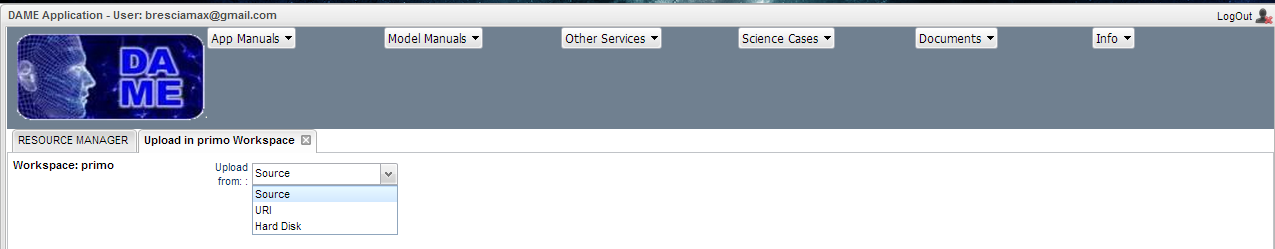}

\caption{The Upload data feature open in a new tab}
\end{figure}
\noindent

\noindent As shown in Fig. 12, when the user selects the ``upload'' command, (label nr. 6 in the Fig. 6), a new tab appears. The user can choose to upload his own data file from, respectively, from any remote URI (a priori known...!) or from his local Hard Disk.

\noindent In the first case (upload from URI), the Fig. 13 shows how to upload a supported type file from a remote address.

\begin{figure}  \centering\includegraphics*[width=\textwidth]{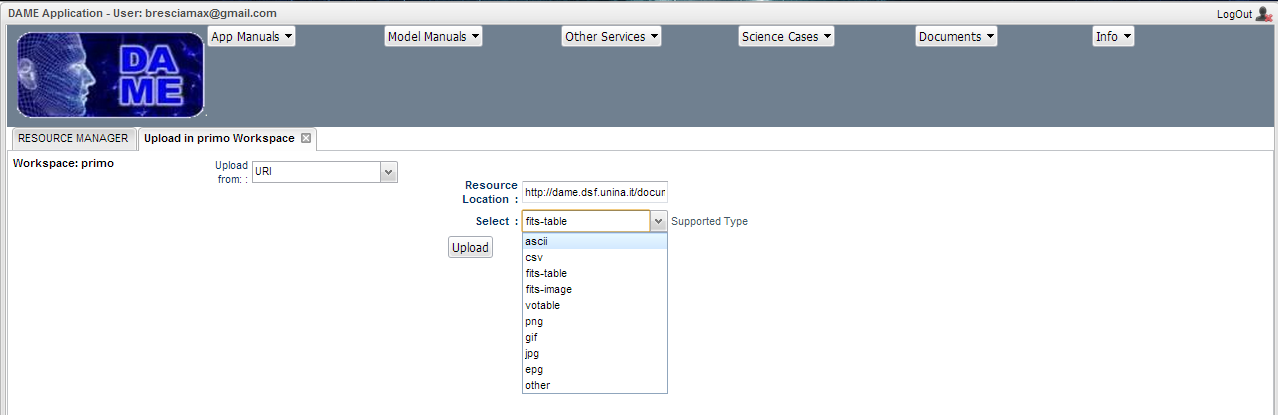}

\caption{ The Upload data from external URI feature}
\end{figure}
\noindent

\noindent In the second case (upload from Hard Disk) the Fig. 14 shows how to select and upload any supported file in the GUI workspace from the user local HD.

\noindent

\begin{figure}  \centering\includegraphics*[width=\textwidth]{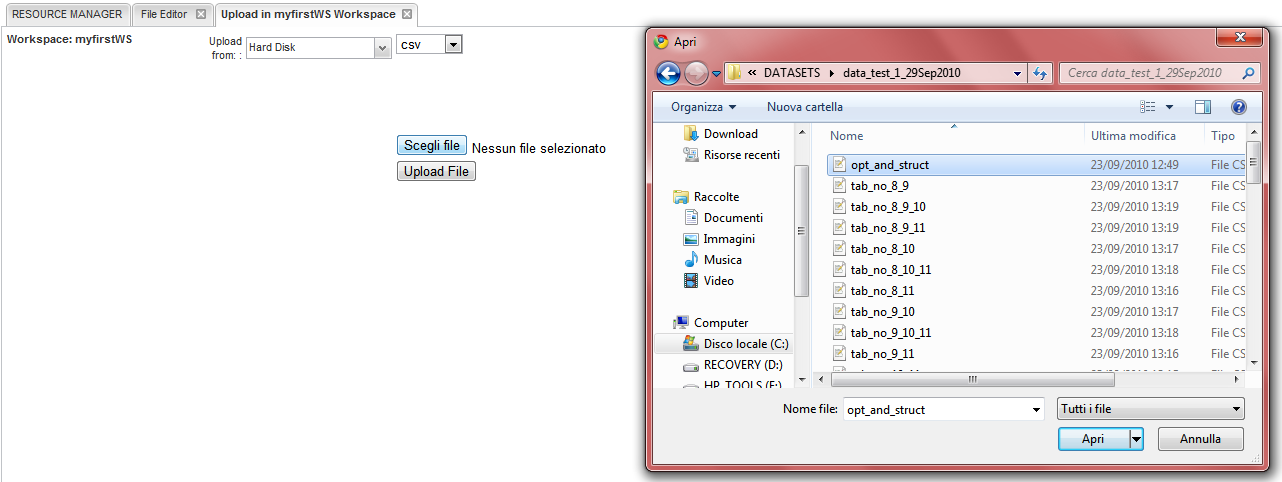}

\caption{ The Upload data from Hard Disk feature}
\end{figure}
\noindent

\noindent After the execution of the operation, coming back to the main GUI tab, the user will found the uploaded file in the ``Files Manager'' sub window related with the currently active workspace, Fig. 15.

\noindent

\begin{figure}  \centering\includegraphics*[width=\textwidth]{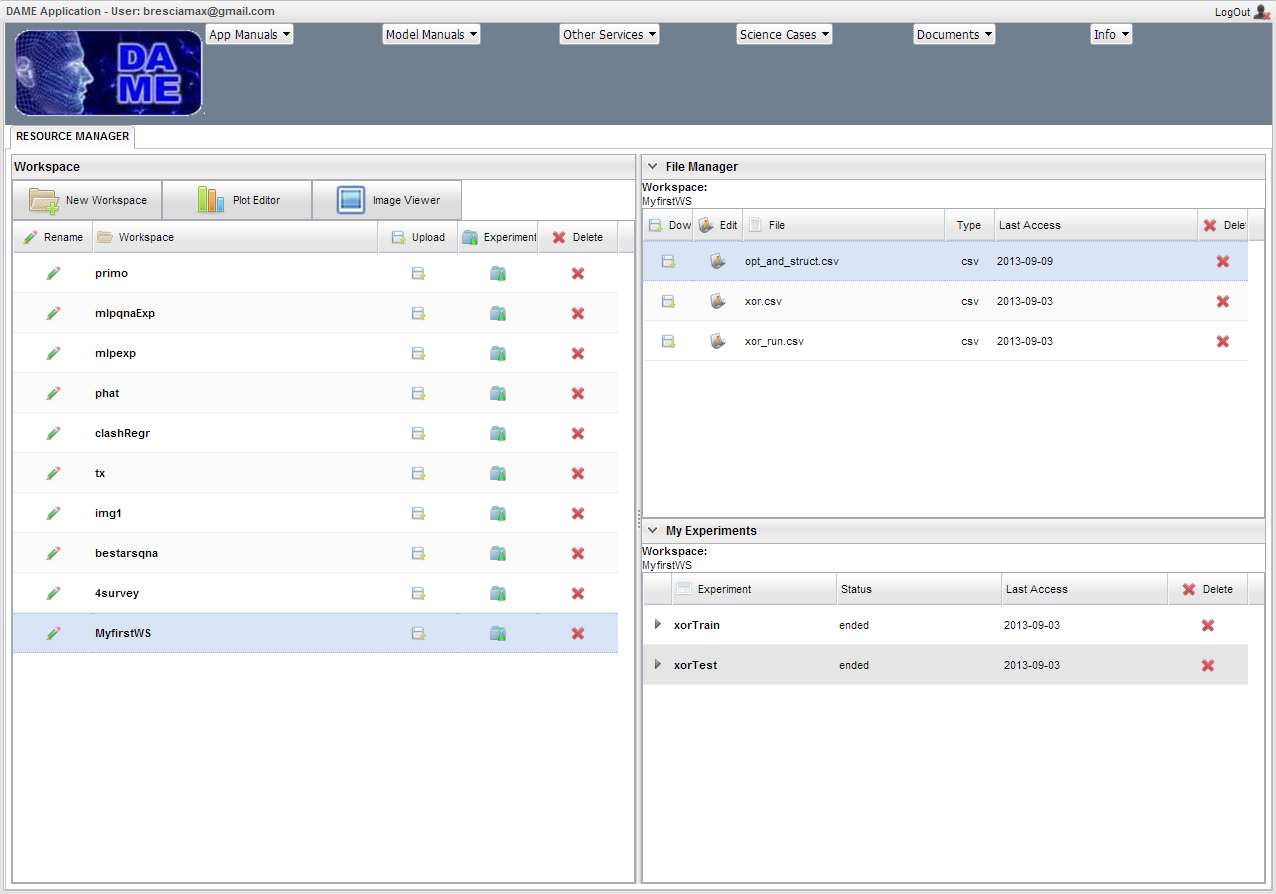}

\caption{ The Uploaded data file in the File Manager sub window}
\end{figure}
\noindent

\paragraph{ How to Create dataset files}

\noindent If the user has already uploaded any supported data file in the workspace, it is possible to select it and to create datasets from it. This is a typical pre-processing phase in a machine learning based experiment, where, starting form an original data file, several different files must be prepared and provided to be submitted as input for, respectively, training, test and validate the algorithm chosen for the experiment. This pre-processing is generally made by applying one or more modification to the original data file (for example obtained from any astronomical observation run or cosmological simulation). The operations available in the web application are the following, Fig. 16:

\noindent

\begin{enumerate}
\item  \textbf{Feature Selection;}

\item \textbf{ Columns Ordering;}

\item \textbf{ Sort Rows by Column;}

\item \textbf{ Column Shuffle;}

\item \textbf{ Row Shuffle;}

\item \textbf{ Split by Rows;}

\item \textbf{ Dataset Scale;}

\item \textbf{ Single Column Scale;}
\end{enumerate}

\noindent

\noindent All these operations, one by one, can be applied starting from a selected data file uploaded in the currently active workspace.

\noindent

\begin{figure}  \centering\includegraphics*[width=4.69in]{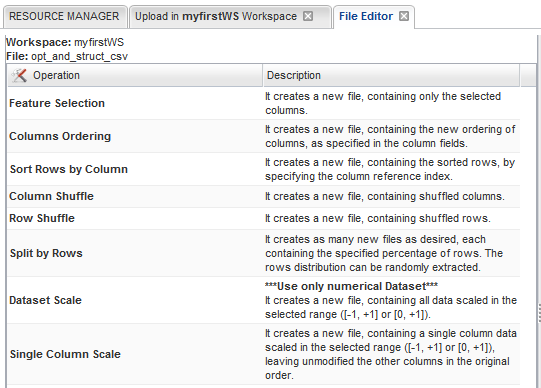}

\caption{ The dataset editor tab with the list of available operations}
\end{figure}

\subparagraph{ Feature Selection}

\noindent This dataset operation permits to select and extract arbitrary number of columns, contained in the original data file, by saving them in a new file (of the same type and with the same extension of the original file), named as columnSubset\_$<$user selected name$>$ (i.e. with specific prefix\textit{columnSubset}). This function is particularly useful to select training columns to be submitted to the algorithm, extracted from the whole data file. Details of the simple procedure are reported in Fig. 17 and Fig. 18.

\noindent

\begin{figure}  \centering\includegraphics*[width=\textwidth]{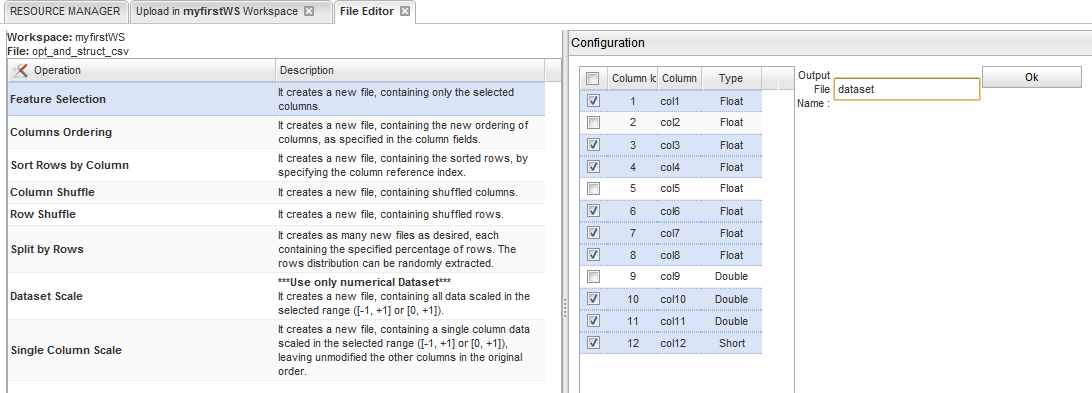}

\caption{ The Feature Selection operation -- select columns and put saving name}
\end{figure}
\noindent

\noindent As clearly visible in Fig. 17, the \textit{Configuration} panel shows the list of columns originally present in the input data file, that can be selected by proper check boxes. Note that the whole content of the data file (in principle a massive data set) is not shown, but simply labelled by column meta-data (as originally present in the file).

\noindent

\noindent

\begin{figure}  \centering\includegraphics*[width=3.29in]{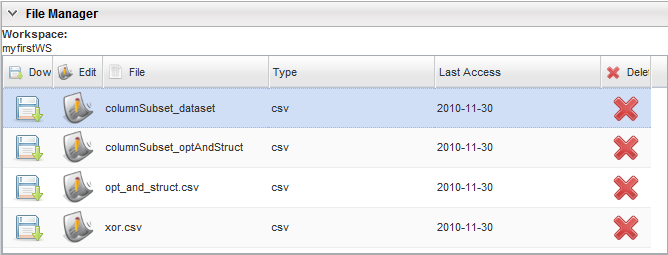}

\caption{The Feature Selection operation -- the new file created}
\end{figure}
\noindent

\subparagraph{ Column Ordering}

\noindent This dataset operation permits to select an arbitrary order of columns, contained in the original data file, by saving them in a new file (of the same type and with the same extension of the original file), named as columnSort\_$<$user selected name$>$ (i.e. with specific prefix\textit{columnSort}). Details of the simple procedure are reported inFig. 20.

\noindent

\noindent

\begin{figure}  \centering\includegraphics*[width=\textwidth]{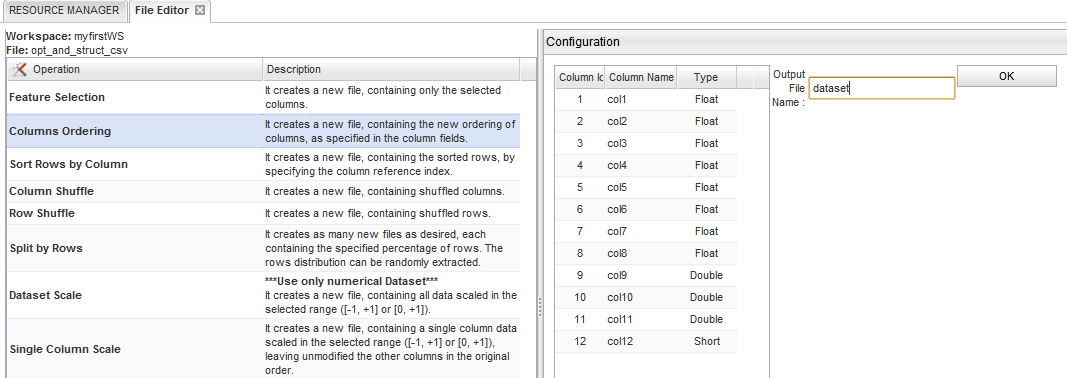}

\caption{The Column Ordering operation -- the starting view}
\end{figure}
\noindent

\noindent In particular, in Fig. 20 it is shown the result of several ``dragging'' operations operated on some columns. By selecting with mouse a column it is possible to drag it in a new desired position . At the end the new saved file will contain the new order given to data columns.

\begin{figure}  \centering\includegraphics*[width=\textwidth]{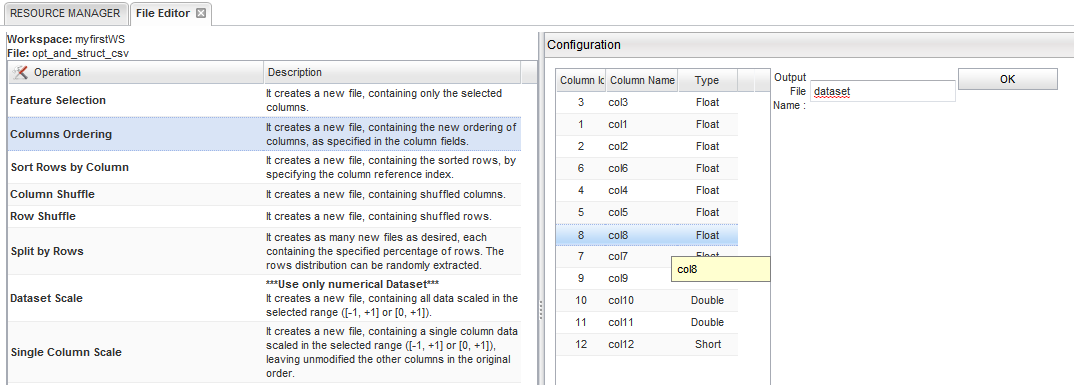}

\caption{The Column Ordering operation -- new order to columns}
\end{figure}
\noindent

\begin{figure}  \centering\includegraphics*[width=3.27in]{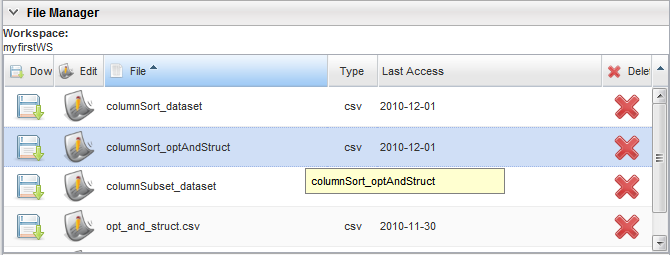}

\caption{ The Column Ordering operation -- new file created}
\end{figure}
\noindent

\subparagraph{ Sort Rows by Column}

\noindent This dataset operation permits to select an arbitrary column, between those contained in the original data file, as sorting reference index for the ordering of all file rows. The result is the creation of a new file (of the same type and with the same extension of the original file), named as rowSort\_$<$user selected name$>$ (i.e. with specific prefix\textit{rowSort}). Details of the simple procedure are reported in Fig. 22, Fig. 23 and Fig. 24.

\noindent

\begin{figure}  \centering\includegraphics*[width=\textwidth]{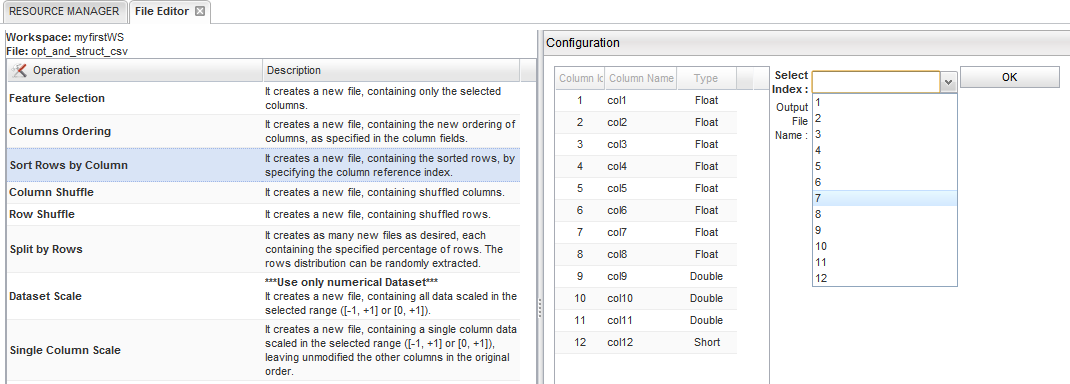}

\caption{The Sort Rows by Column operation -- step 1}
\end{figure}
\noindent

\noindent

\begin{figure}  \centering\includegraphics*[width=\textwidth]{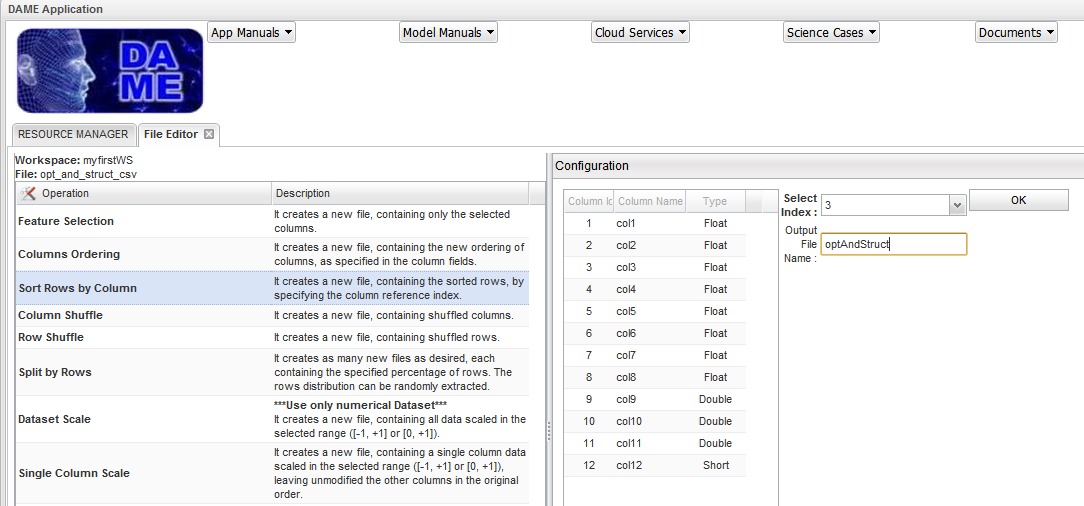}

\caption{The Sort Rows by Column operation -- step 2}
\end{figure}
\noindent

\noindent

\begin{figure}  \centering\includegraphics*[width=3.03in]{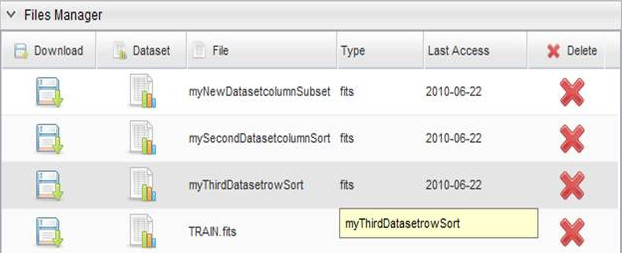}

\caption{The Sort Rows by Column operation -- the new file created}
\end{figure}
\noindent

\noindent

\subparagraph{ Column Shuffle}

\noindent This dataset operation permits to operate a random shuffle of the columns, contained in the original data file. The result is the creation of a new file (of the same type and with the same extension of the original file), named as shuffle\_$<$user selected name$>$ (i.e. with specific prefix\textit{shuffle}). Details of the simple procedure are reported in Fig. 25 and Fig. 26.

\noindent

\begin{figure}  \centering\includegraphics*[width=\textwidth]{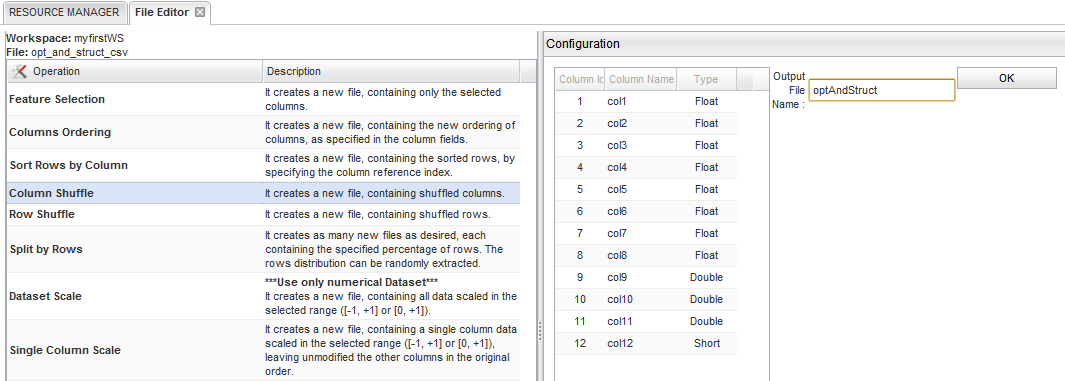}

\caption{The Column Shuffle operation -- step 1}
\end{figure}
\noindent

\noindent

\begin{figure}  \centering\includegraphics*[width=3.27in]{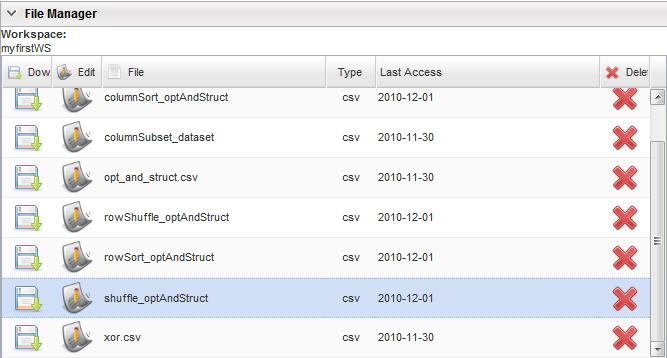}

\caption{The Column Shuffle operation -- the new file created}
\end{figure}
\noindent

\subparagraph{ Row Shuffle}

\noindent This dataset operation permits to operate a random shuffle of the rows, contained in the original data file. The result is the creation of a new file (of the same type and with the same extension of the original file), named as rowShuffle\_$<$user selected name$>$ (i.e. with specific prefix\textit{rowShuffle}). Details of the simple procedure are reported in Fig. 27 and Fig. 28.

\noindent

\begin{figure}  \centering\includegraphics*[width=\textwidth]{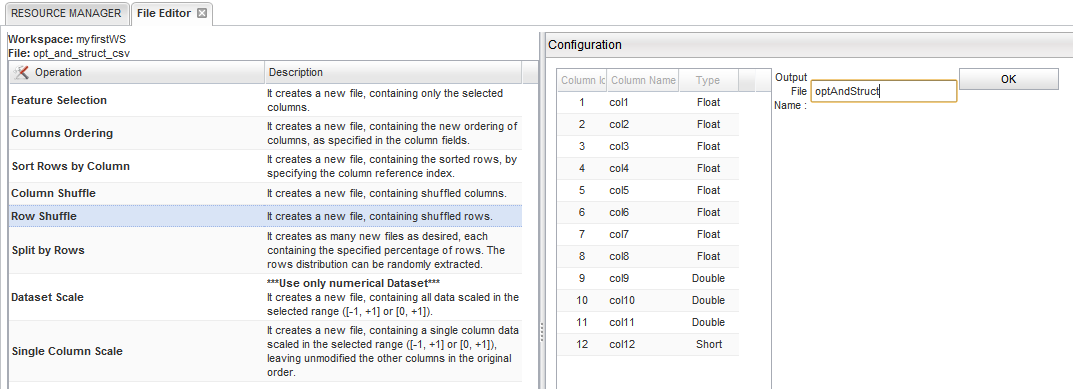}

\caption{The Row Shuffle operation -- step 1}
\end{figure}
\noindent

\noindent

\begin{figure}  \centering\includegraphics*[width=3.29in]{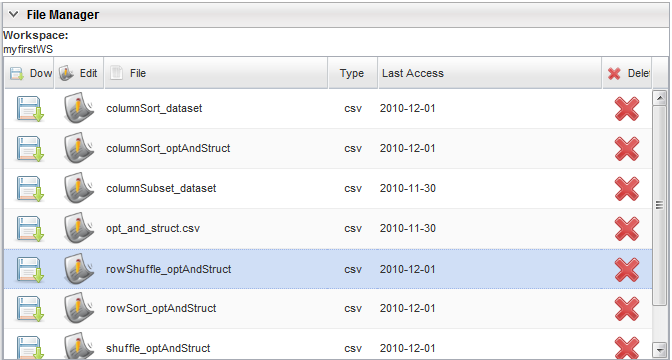}

\caption{The Row Shuffle operation -- the new file created}
\end{figure}
\noindent

\subparagraph{ Split by Rows}

\noindent This dataset operation permits to split the original file into two new files containing the selected percentages of rows, as indicated by the user. The user can move one of the two sliding bars in order to fix the desired percentage. The other sliding bar will automatically move in the right percentage position. The new file names are those filled in by the user in the proper name fields as split1\_$<$user selected name$>$(split2\_$<$user selected name$>$) (i.e. with specific prefix\textit{split1}and \textit{split2}). Details of the simple procedure are reported inFig. 29, Fig. 30.

\noindent

\begin{figure}  \centering\includegraphics*[width=\textwidth]{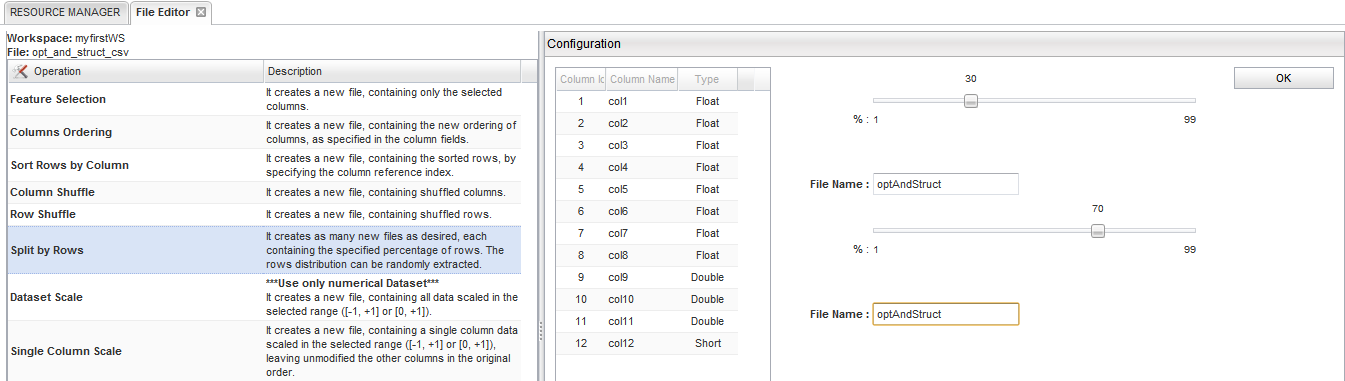}

\caption{The Split by Rows operation -- step 1}
\end{figure}
\noindent

\noindent

\begin{figure}  \centering\includegraphics*[width=3.29in]{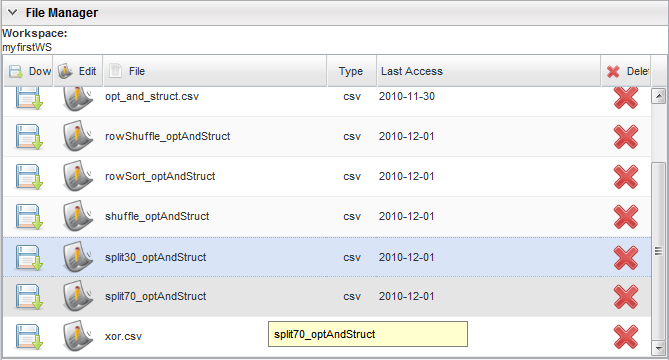}

\caption{The Split by Rows operation -- the new files created}
\end{figure}
\noindent

\subparagraph{ Dataset Scale}

\noindent This dataset operation (that works on numerical data files only!)  permits to normalize column data in one of two possible ranges, respectively, [-1, +1] or [0, +1]. This is particularly frequent in machine learning experiments to submit normalized data, in order to achieve a correct training of internal patterns. The result is the creation of a new file (of the same type and with the same extension of the original file), named as scale\_$<$user selected name$>$ (i.e. with specific prefix\textit{scale}). Details are reported in Fig. 31.

\noindent

\begin{figure}  \centering\includegraphics*[width=\textwidth]{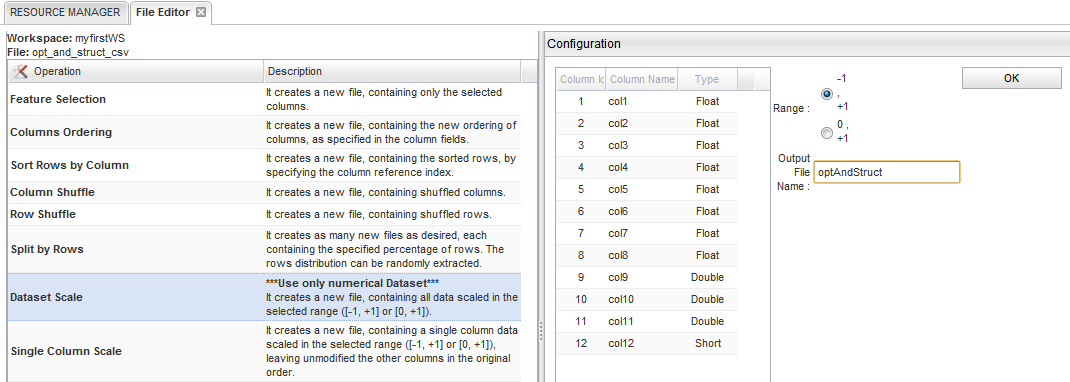}

\caption{The Dataset Scale operation -- step 1}
\end{figure}
\noindent

\subparagraph{ Single Column  Scale}

\noindent This dataset operation (that works on numerical data files only!)  permits to normalize a single selected column, between those contained in the original file, in one of two possible ranges, respectively, [-1, +1] or [0, +1]. The result is the creation of a new file (of the same type and with the same extension of the original file), named as scaleOneCol\_$<$user selected name$>$ (i.e. with specific prefix\textit{scaleOneCol}). Details of the simple procedure are reported in Fig. 32.

\noindent

\begin{figure}  \centering\includegraphics*[width=\textwidth]{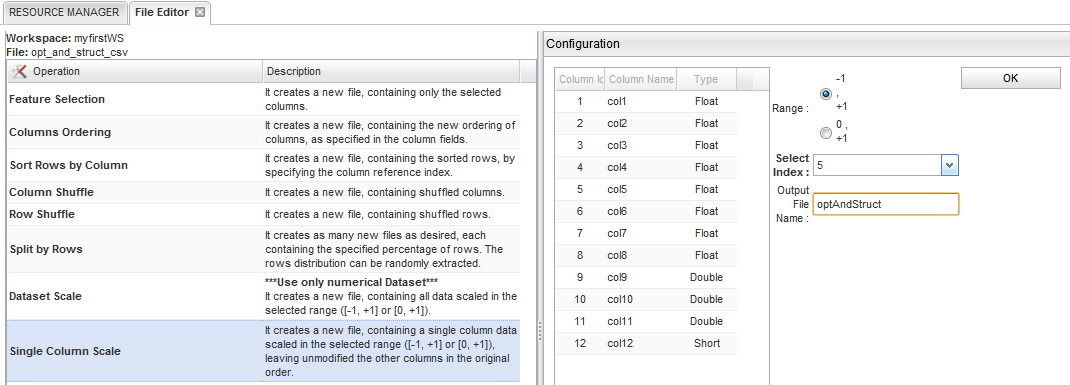}

\caption{The Single Column Scale operation -- step 1}
\end{figure}
\noindent

\paragraph{ Download data}

\noindent All data files (not only those of supported type) listed in the workspace and/or in the experiment panels, respectively, ``Files Manager'' and ``Experiment Manager'', can be downloaded by the user on his own hard disk, by simply selecting the icon labelled with ``Download'' in the mentioned panels.

\paragraph{ Moving data files}

\noindent The virtual separation of user data files between workspace and experiment files, located in the respective panels (``File Manager'' for workspace files, and ``My Experiments'' for experiment files), is due to the different origin of such files and depends on their registration policy into the web application database. The data files present in the workspace list (``File Manager'' area panel) are usually registered as ``input'' files, i.e. to be submitted as inputs for experiments. While others, present in the experiment list (``My Experiments'' panel), are considered as ``output'' files, i.e. generated by the web application after the execution of an experiment.

\noindent It is not rare, in machine learning complex workflows, to re-use some output files, obtained after training phase, as inputs of a test/validation phase of the same workflow. This is true for example for a MLP weight matrix file, output of the training phase, to be re-used as input weight matrix of a test (or validation) session of the same network.

\noindent In order to make available this fundamental feature in our application, the icon command nr. 18 (AddInWS) in Fig. 6, associated to each output file of an experiment, can be selected by the user in order to ``copy'' the file from experiment output list to the workspace input list, becoming immediately available as input file for new experiments belonging to the \textbf{\underbar{same}} workspace:\textbf{: as important remark, in the beta release it is not yet possible to ``move'' files from a workspace to another}. The alternative procedure to perform this action is to download the file on user local Hard Disk and to upload it into another desired workspace in the webapp.

\subsection{ Plotting and Visualization}

\noindent The final release of the web application offers two new options for plotting and visualization of data (tables or images).

\paragraph{ Plotting}

\noindent By pressing the ``Plot Editor'' button in the main menu a series of plot tabs will appear. Each one is dedicated to a specific type of plot, for instance Histogram, Scatter Plot 2D, Scatter Plot 3D and Line Plot.

\noindent

\begin{figure}  \centering\includegraphics*[width=\textwidth]{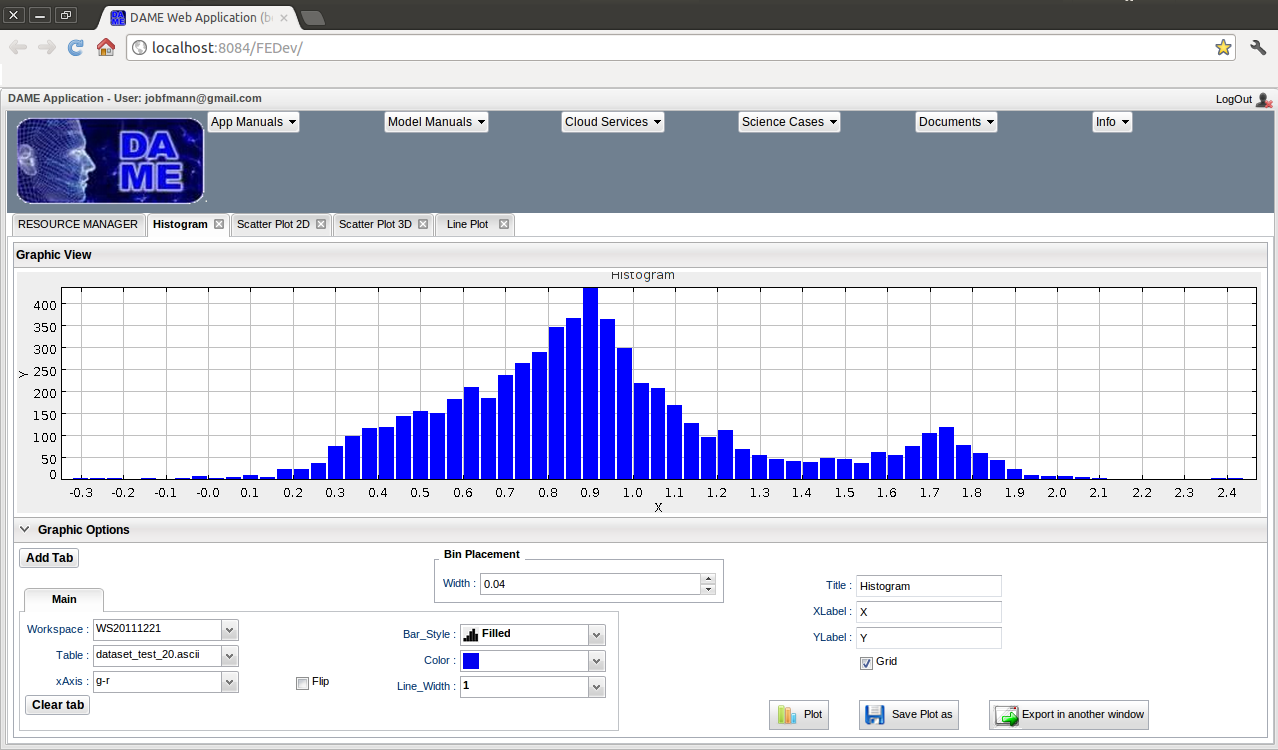}

\caption{ The Histogram tab}
\end{figure}
\noindent As shown in Fig. 33 there is the possibility to create and visualize an histogram of any table file previously loaded or produced in the web application.

\noindent There are several options:

\begin{enumerate}
\item  Workspace: the user workspace hosting the table;

\item  Table: the name of the table to be plotted;

\item  xAxis: selection of the column of table to be plotted;

\item  Bar\_Style: style of the bars;

\item  Color: color of the plot bars;

\item  Line\_Width: width of the bars;

\item  Flip: enable the flipping of the x Axis of the plot;

\item  Title: title of the plot;

\item  Xlabel: label of the x axis;

\item  Ylabel: label of the y axis;

\item  Grid: enable/disable the grid in the plot;

\item  Bin Placement: change the bin width of plot;

\item  Clear Tab: clear the tab;

\item  Plot: creation and visualization of the selected histogram;

\item  Save Plot As: plot saving with user typed name;

\item  Export in another window: the plot will be moved in an independent tab of the web browser;

\item  Add Tab: enable the creation of a multi layer histogram as shown in Fig. 34.
\end{enumerate}

\noindent

\begin{figure}  \centering\includegraphics*[width=\textwidth]{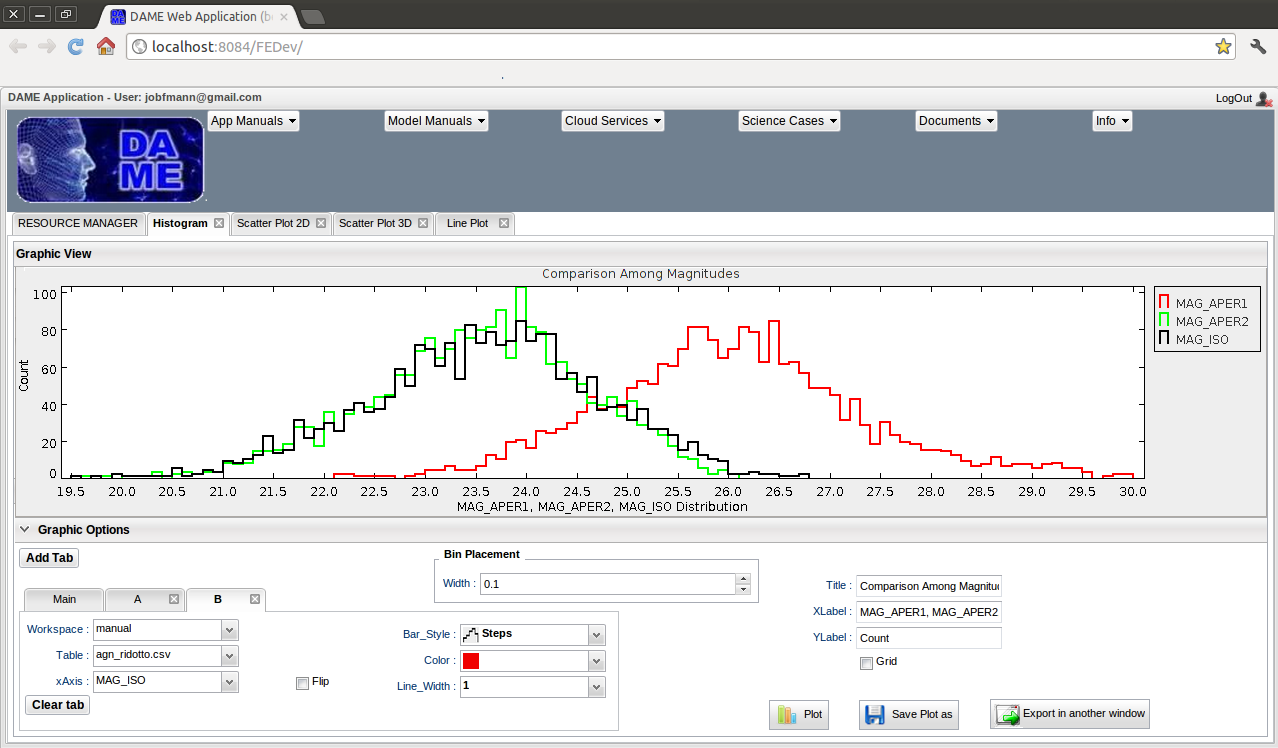}

\caption{ A multi layer histogram plot}
\end{figure}
\noindent

\noindent \textbf{ADVERTISEMENT:} whenever the user change any parameter of the current plot, it is needed to click the button ``Plot'' to refresh the visualized plot.

\noindent

\begin{figure}  \centering\includegraphics*[width=\textwidth]{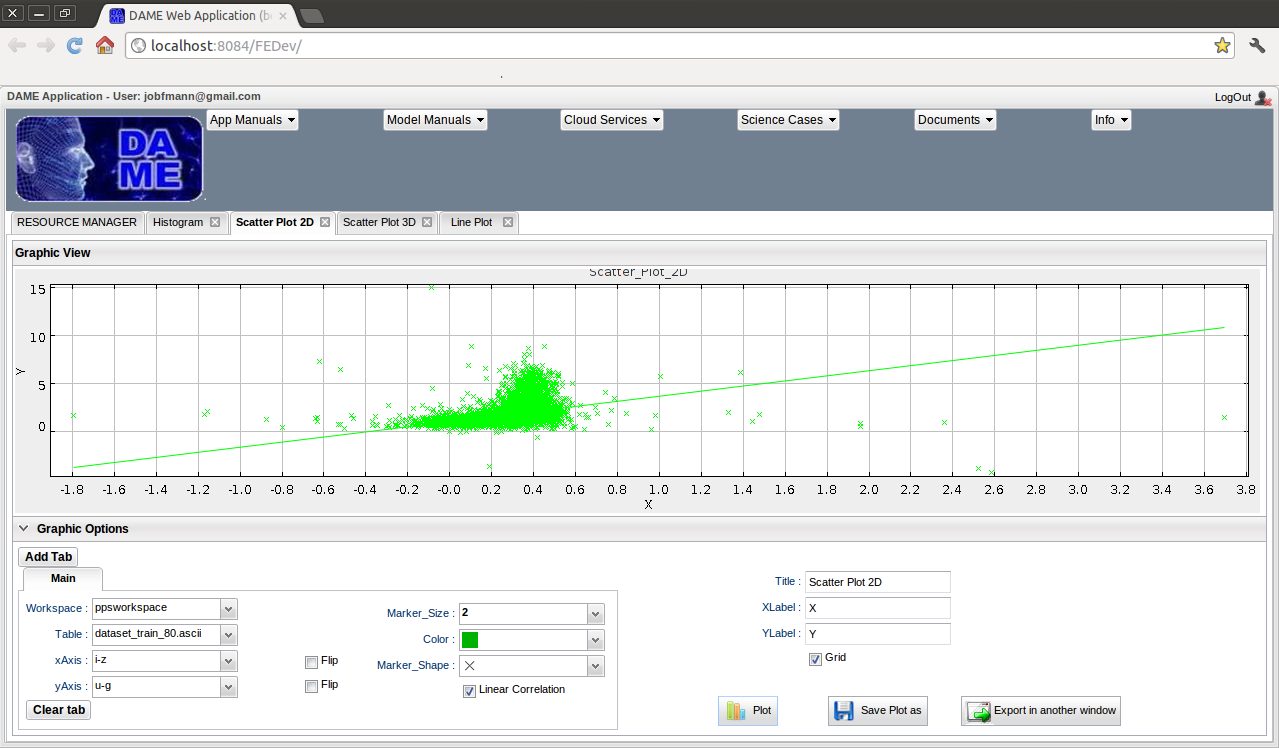}

\caption{ The Scatter 2D tab}
\end{figure}
\noindent As shown in Fig. 35 there is the possibility to create and visualize a scatter 2D of any table file previously loaded or produced in the web application.

\noindent There are several options:

\begin{enumerate}
\item  Workspace: the user workspace hosting the table;

\item  Table: the name of the table to be plotted;

\item  xAxis: selection of the x column of table to be plotted;

\item  yAxis: selection of the y column of table to be plotted;

\item  Marker\_Size: size of the marker;

\item  Color: color of the plot bars;

\item  Marker\_Shape: shape of the marker;

\item  Line\_Width: width of the bars;

\item  Linear Correlation: enable the drawing of a line based on linear correlation of columns;

\item  Flip: enable the flipping of the x Axis of the plot;

\item  Flip: enable the flipping of the y Axis of the plot;

\item  Title: title of the plot;

\item  Xlabel: label of the x axis;

\item  Ylabel: label of the y axis;

\item  Grid: enable/disable the grid in the plot;

\item  Clear Tab: clear the tab;

\item  Plot: creation and visualization of the selected histogram;

\item  Save Plot As: plot saving with user typed name;

\item  Export in another window: the plot will be moved in an independent tab of the web browser;

\item  Add Tab: enable the creation of a multi layer scatter 2D as shown in Fig. 36.
\end{enumerate}

\noindent

\noindent

\begin{figure}  \centering\includegraphics*[width=\textwidth]{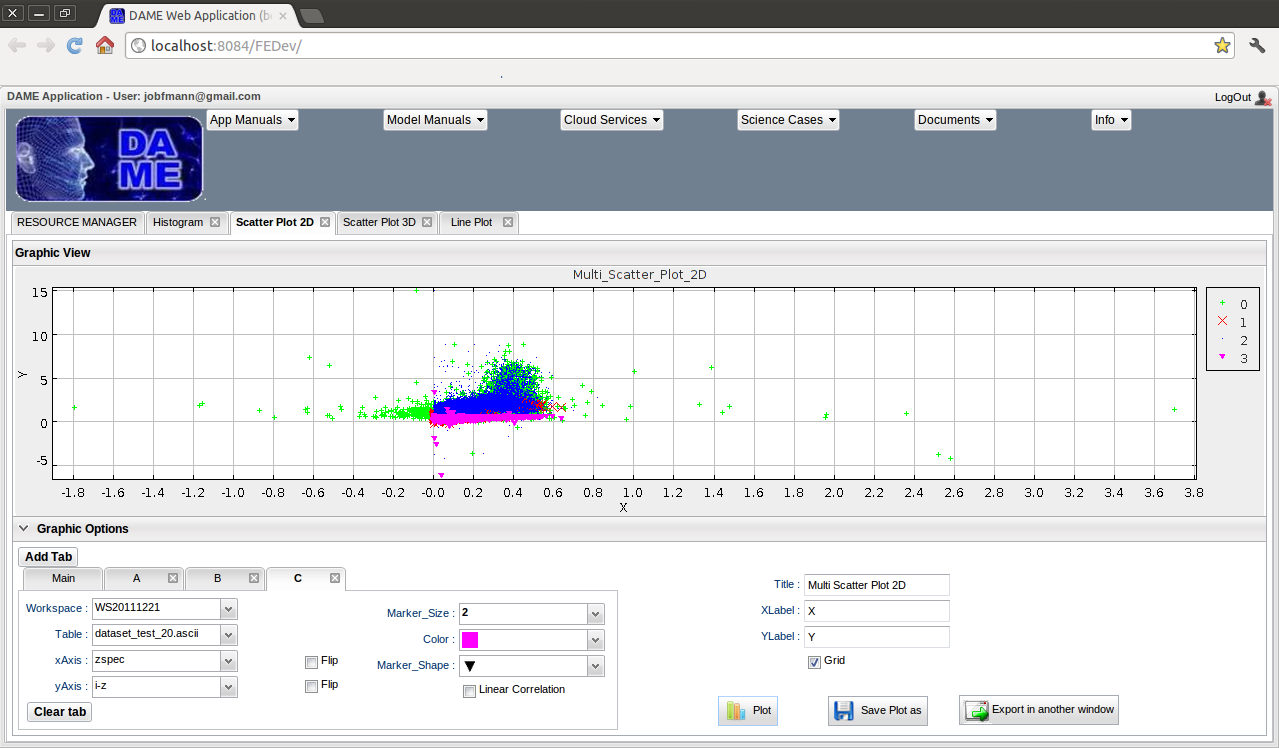}

\caption{ A multi layer scatter 2D plot}
\end{figure}
\noindent \textbf{ADVERTISEMENT:} whenever the user change any parameter of the current plot, it is needed to click the button ``Plot'' to refresh the visualized plot.

\noindent

\noindent

\begin{figure}  \centering\includegraphics*[width=\textwidth]{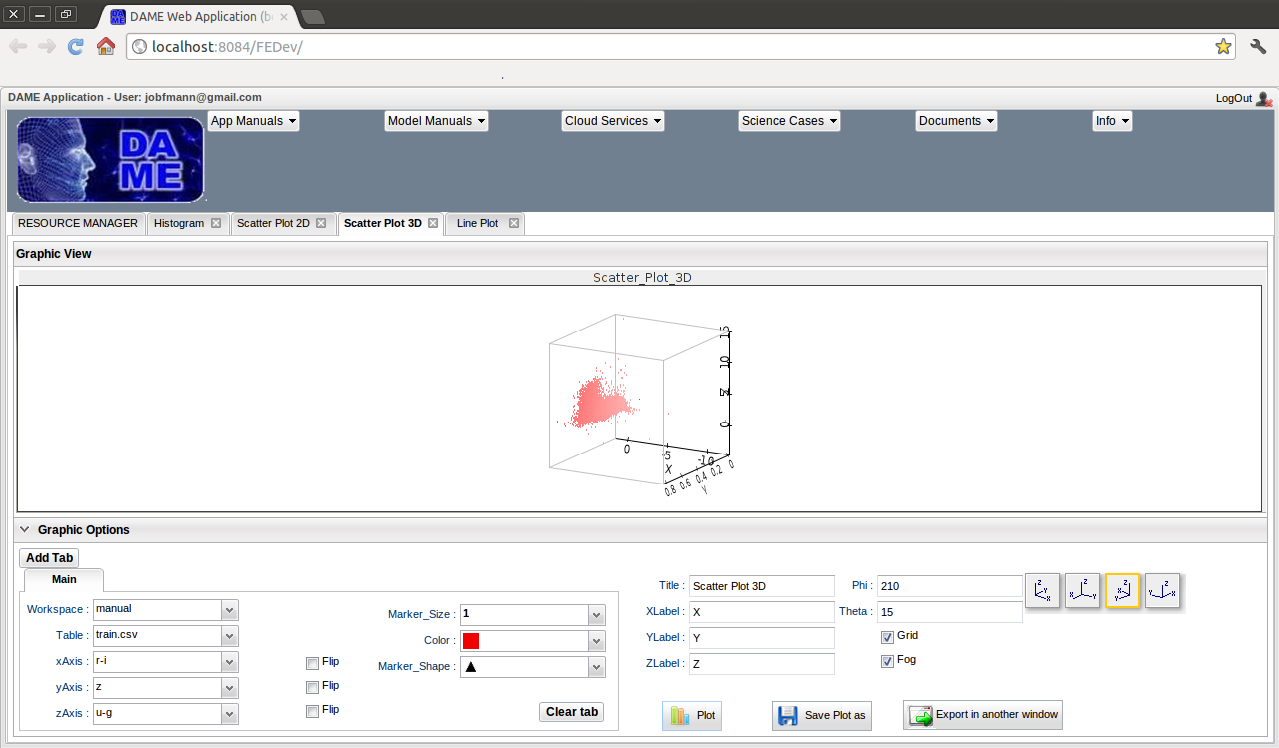}

\caption{ The Scatter Plot 3D tab}
\end{figure}
\noindent As shown in Fig. 37 there is the possibility to create and visualize a scatter 3D of any table file previously loaded or produced in the web application.

\noindent There are several options:

\begin{enumerate}
\item  Workspace: the user workspace hosting the table;

\item  Table: the name of the table to be plotted;

\item  xAxis: selection of the x column of table to be plotted;

\item  yAxis: selection of the y column of table to be plotted;

\item  zAxis: selection of the z column of table to be plotted;

\item  Marker\_Size: size of the marker;

\item  Color: color of the plot bars;

\item  Marker\_Shape: shape of the marker;

\item  Line\_Width: width of the bars;

\item  Flip: enable the flipping of the x Axis of the plot;

\item  Flip: enable the flipping of the y Axis of the plot;

\item  Flip: enable the flipping of the z Axis of the plot;

\item  Title: title of the plot;

\item  Xlabel: label of the x axis;

\item  Ylabel: label of the y axis;

\item  Zlabel: label of the z axis;

\item  Grid: enable/disable the grid in the plot;

\item  Fog: enable/disable the fog effect;

\item  Phi: rotation angle in degrees;

\item  Theta: rotation angle in degrees;

\item  Orientation buttons: four predefined couples of Phi and Theta;

\item  Clear Tab: clear the tab;

\item  Plot: creation and visualization of the selected histogram;

\item  Save Plot As: plot saving with user typed name;

\item  Export in another window: the plot will be moved in an independent tab of the web browser;

\item  Add Tab: enable the creation of a multi layer scatter plot 3D.
\end{enumerate}

\noindent

\noindent \textbf{ADVERTISEMENT:} whenever the user change any parameter of the current plot, it is needed to click the button ``Plot'' to refresh the visualized plot.

\noindent

\begin{figure}  \centering\includegraphics*[width=\textwidth]{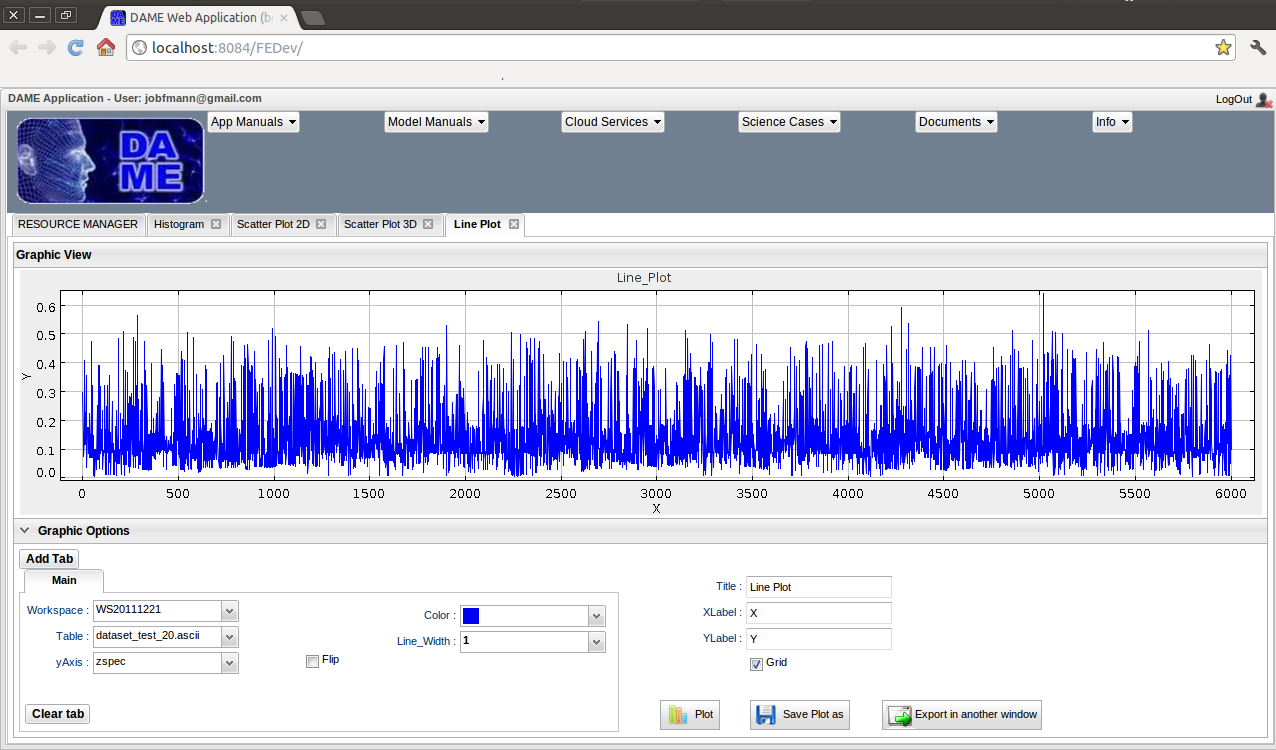}

\caption{ The Line Plot tab}
\end{figure}
\noindent As shown in Fig. 38 there is the possibility to create and visualize an histogram of any table file previously loaded or produced in the web application.

\noindent There are several options:

\begin{enumerate}
\item  Workspace: the user workspace hosting the table;

\item  Table: the name of the table to be plotted;

\item  yAxis: selection of the column of table to be plotted;

\item  Color: color of the plot line;

\item  Line\_Width: width of the line;

\item  Flip: enable the flipping of the x Axis of the plot;

\item  Title: title of the plot;

\item  Xlabel: label of the x axis;

\item  Ylabel: label of the y axis;

\item  Grid: enable/disable the grid in the plot;

\item  Clear Tab: clear the tab;

\item  Plot: creation and visualization of the selected histogram;

\item  Save Plot As: plot saving with user typed name;

\item  Export in another window: the plot will be moved in an independent tab of the web browser;

\item  Add Tab: enable the creation of a multi layer line plot as shown in Fig. 39.
\end{enumerate}

\noindent

\noindent

\begin{figure}  \centering\includegraphics*[width=\textwidth]{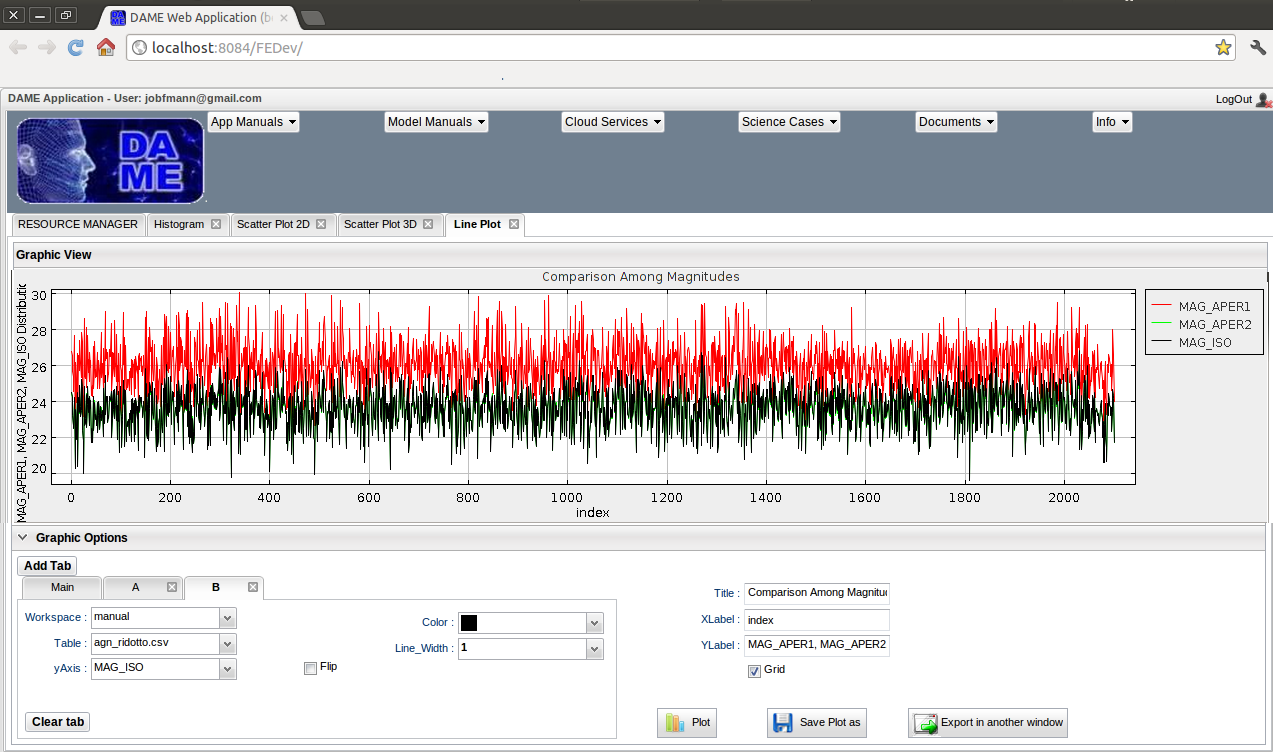}

\caption{ A multi layer line plot}
\end{figure}
\noindent \textbf{ADVERTISEMENT:} whenever the user change any parameter of the current plot, it is needed to click the button ``Plot'' to refresh the visualized plot.

\noindent

\paragraph{ Visualization}

\noindent This option can be enabled from the main tab of the GUI by simply clicking on the menu button ``Image Viewer''. A dedicated tab will appear in the Resource Manager giving the possibility to load and visualize any image previously uploaded or produced in the web application.

\noindent

\begin{figure}  \centering\includegraphics*[width=\textwidth]{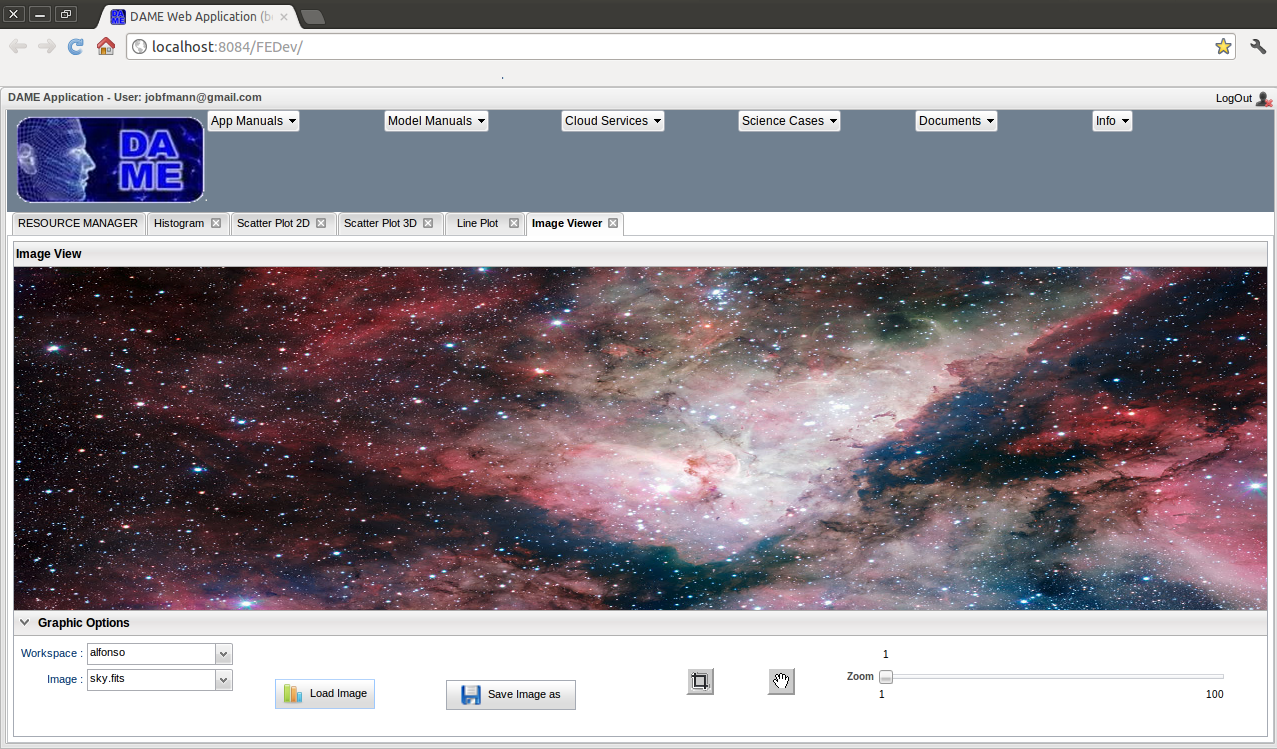}

\caption{ The visualization tab}
\end{figure}
\noindent As shown in Fig. 40 the visualization tab offer the following options:

\begin{enumerate}
\item  Workspace: the user workspace hosting the image;

\item  Image: the name of the image to be visualized;

\item  Load Image: after the selection of workspace and image this button shows the image;

\item  Crop: button used to crop the image;

\item  Hand: button used to move the image;

\item  Zoom: sliding bar used to zoom the image;

\item  Save Image as: button used to save the modified image.
\end{enumerate}

\noindent

\noindent \textbf{ADVERTISEMENT:} multi image fits files are not supported by this functionality.

\noindent

\subsection{ Experiment Management}

\noindent After creating at least one workspace, populating it with input data files (of supported type) and optionally creatingany dataset file, the next logical operation required is the configuration and launch of an experiment.

\noindent In what follows, we will explain the experiment configuration and execution by making use of an example (very simple not linearly separable XOR problem) which can be replicated by the user by using the xor.csv and xor\_run.csv data files (downloadable from the beta intro web page,

\noindent http://dame.dsf.unina.it/dameware.html ).

\noindent

\noindent The Fig. 41 shows the initial step required, i.e. the selection of the icon command nr. 7of Fig. 6 in order to create the new experiment.

\noindent

\begin{figure}  \centering\includegraphics*[width=\textwidth]{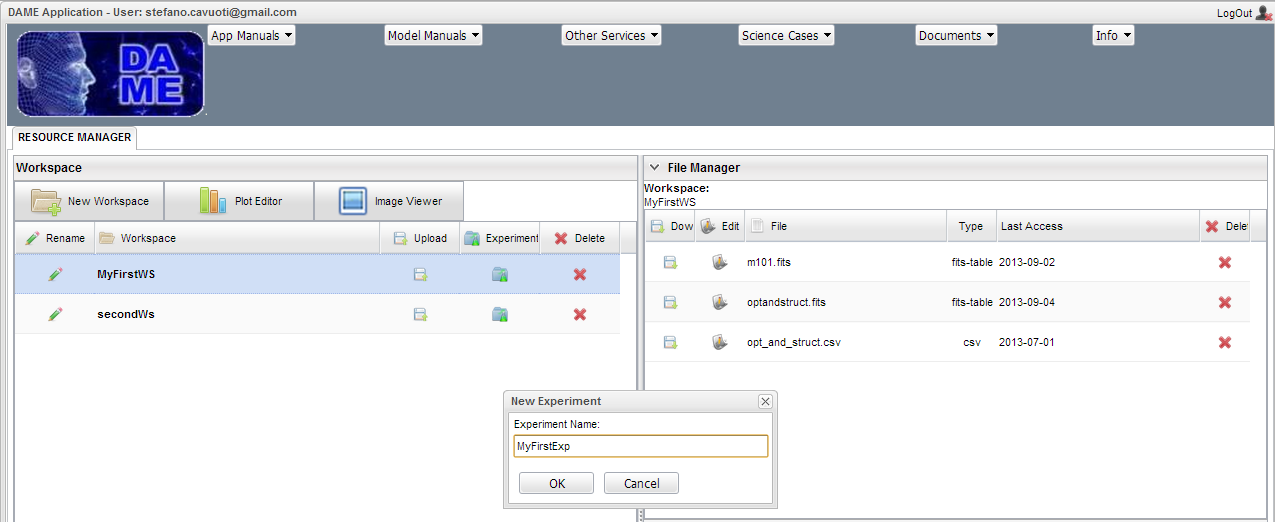}

\caption{ Creating a new experiment (by selecting icon ``Experiment'' in the workspace)}
\end{figure}
\noindent

\noindent Immediately after, an automatic new tab appears, making available all basic features to select, configure and launch the experiment. In particular there is the list of couples [functionality]-[model] to choose for the current experiment. The proper choice should be done in order to solve a particular problem. It depends basically on the dataset to be used as input and on the output the user wants to obtain. Please, refer to the particular model reference manual for more details.

\noindent

\begin{figure}  \centering\includegraphics*[width=2.67in]{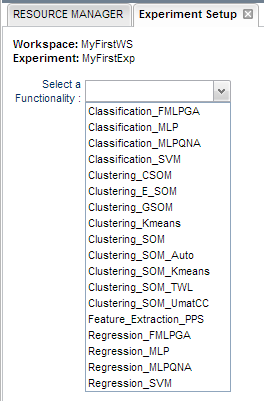}

\caption{ The new tab open after creation of a new experiment with the list of available options}
\end{figure}
\noindent

\noindent The user can choose between classification, regression or clustering type of functionality to be applied to his problem. Each of these functionalities can be achieved by associating a particular data mining model, chosen between following types:

\noindent

\begin{enumerate}
\item  \textbf{MLP} : Multilayer Perceptron \cite{23} neural network trained by standard Back Propagation (descent gradient of the error) learning rule. Associated functionalities are classification and regression;

\item  \textbf{FMLPGA}: Fast Multilayer Perceptron neural network trained by Genetic Algorithm \cite{18} learning rule. Associated functionalities are classification and regression. This model is available in two versions: CPU and GPU; the second one is the parallelized version;

\item  \textbf{SVM}: Support Vector Machine \cite{13} model. Associated functionalities are classification and regression;

\item  \textbf{MLPQNA}: Multilayer Perceptron neural network trained by Quasi Newton learning rule (Cavuoti et al. 2012). Associated functionalities are classification and regression;

\item  \textbf{LEMON}: Multilayer Perceptron neural network trained by Levenberg-Marquardt learning rule \cite{19}. Associated functionalities are classification and regression;

\item  \textbf{RANDOM FOREST}: Randomly generated forest of decision trees network \cite{1}. Associated functionalities are classification and regression;

\item  \textbf{KMEANS}: Standard Kmeans algorithm \cite{21}. Associated functionality is clustering;

\item  \textbf{CSOM}: Customized Self Organizing Feature Map (SOFM, \cite{20}) for clustering on FITS images;

\item  \textbf{GSOM}: Gated Self Organizing Map (SOFM, \cite{20}) for clustering on text and/or image files;

\item  \textbf{PPS:} Probabilistic Principal Surfaces for feature extraction;\textbf{}

\item \textbf{ SOM: }Self organizing Map  \cite{20} for pre-clustering on text or image files;\textbf{}

\item \textbf{ SOM + Auto: }SOM with an automatized post processing phase for clustering on text or image files \cite{17};\textbf{}

\item \textbf{ SOM + Kmeans: }SOM with a Kmeans based post processing phase for clustering on text or image files \cite{17};\textbf{}

\item \textbf{ SOM + TWL: }SOM with a Two Winners Linkage (TWL) based post processing phase for clustering on text or image files \cite{17};\textbf{}

\item \textbf{ SOM + UmatCC: }SOM with an U-matrix Connected Components (UmatCC) based post processing phase for clustering on text or image files \cite{17};\textbf{}

\item \textbf{ ESOM: }Evolving SOM \cite{15} for pre-clustering on text or image files.\textbf{}

\item \textbf{ STATISTICS: }tool which derives usefull statistics for regression experiments\textbf{}
\end{enumerate}

\noindent

\noindent Specific related manuals are available to obtain detailed information about the use of the above models (see webapp header menu options).

\noindent

\noindent After the selection of the proper functionality-model, the tab will show (greyed) some options and the possibility to select the use case. The greyed options (like help button) will be activated after the selection of the use case to be configured and launched.

\noindent

\begin{figure}  \centering\includegraphics*[width=3.58in]{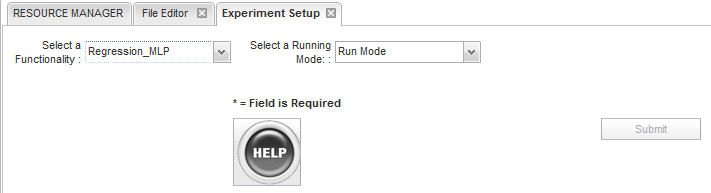}

\caption{ The new state of the experiment configuration tab after the selection of the model}
\end{figure}
\noindent

\noindent As known, data mining models, following machine learning paradigm, offer a series of use cases (see figures below):

\noindent

\begin{enumerate}
\item  \textbf{\underbar{Train}}: training (learning) phase in which the model is trained with the user available BoK;
\end{enumerate}

\noindent

\begin{figure}  \centering\includegraphics*[width=3.48in]{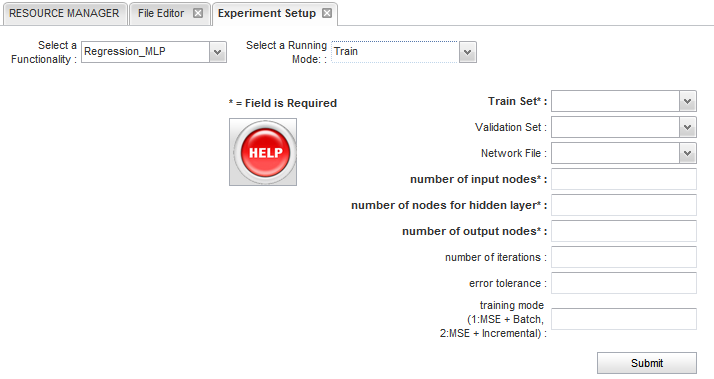}

\caption{ The configuration options in the Train use case}
\end{figure}
\noindent

\begin{enumerate}
\item  \textbf{\underbar{Test}}: a sort of validation of the training phase. It can done by submitting the same training dataset, or a subset or a mix between already submitted and new dataset patterns;
\end{enumerate}

\noindent

\begin{figure}  \centering\includegraphics*[width=3.46in]{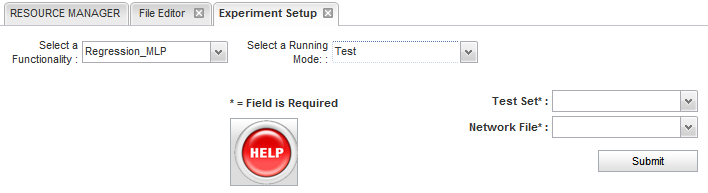}

\caption{ The configuration options in the Test use case}
\end{figure}
\noindent

\begin{enumerate}
\item  \textbf{\underbar{Run}}: normal use of the already trained model;
\end{enumerate}

\noindent

\begin{figure}  \centering\includegraphics*[width=3.46in]{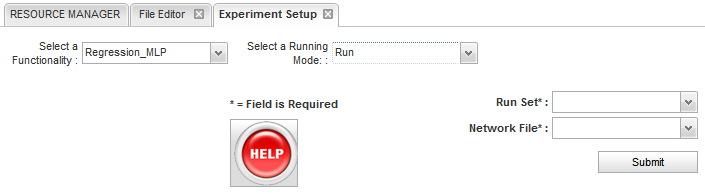}

\caption{ The configuration options in the Run use case}
\end{figure}
\noindent

\begin{enumerate}
\item  \textbf{\underbar{Full}}: the complete and automatic serialized execution of the three previous use cases (train, test and Run). It is a sort of workflow, considered as a complete and exhaustive experiment for a specific problem.
\end{enumerate}

\noindent

\begin{figure}  \centering\includegraphics*[width=3.48in]{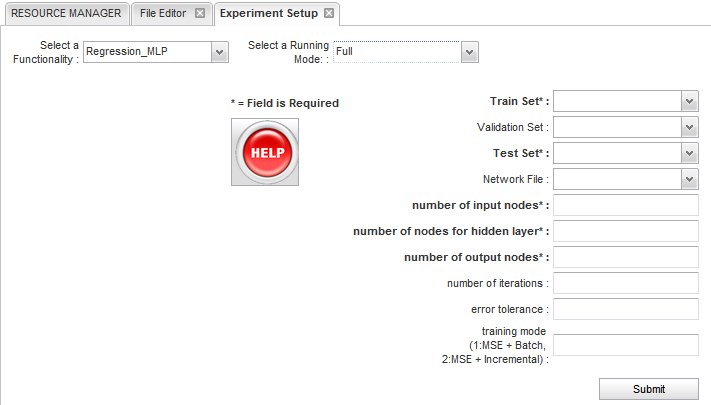}

\caption{ The configuration options in the Full use case}
\end{figure}
\noindent

\noindent In all the above use case tabs, the help button redirects to a specific web page, reporting in verbose mode detailed description of all parameters. In particular, the parameter fields marked by an asterisk are considered ``required'' by the user. All other parameters can be left empty, by assuming a default value (also reported in the hep page).

\noindent

\begin{figure}  \centering\includegraphics*[width=\textwidth]{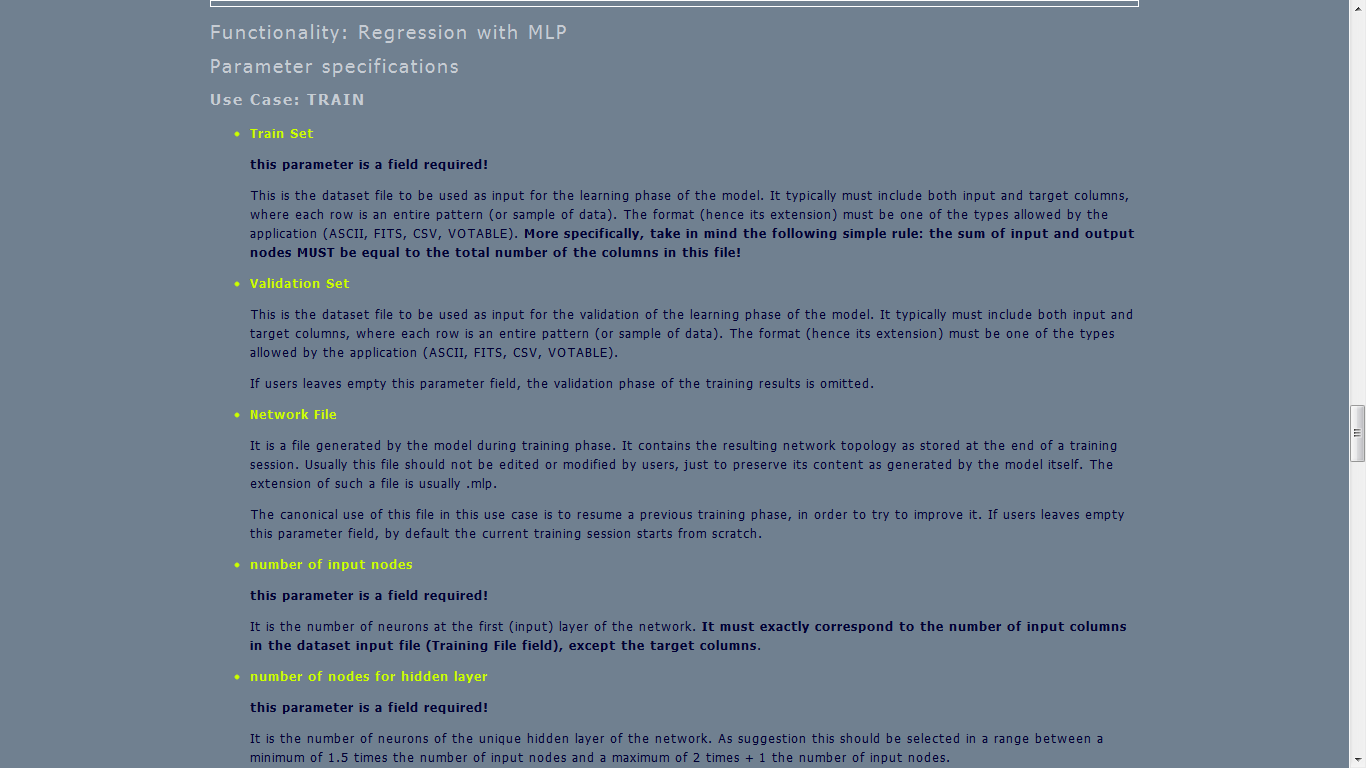}

\caption{ Example of a web page automatically open after the click on the help button}
\end{figure}
\noindent

\noindent After completion of the parameter configuration, the ``Submit'' button launches the experiment.

\noindent

\noindent After launch of an experiment, it can result in one of the following states:

\noindent

\begin{enumerate}
\item  \textbf{Enqueued}: the execution is put in the job queue;

\item  \textbf{Running}: the experiment has been launched and it is running;

\item  \textbf{Failed}: the experiment has been stopped or concluded with any error occurred;

\item  \textbf{Ended}: the experiment has been successfully concluded;
\end{enumerate}

\noindent

\noindent

\begin{figure}  \centering\includegraphics*[width=\textwidth]{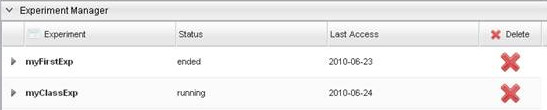}

\caption{ Some different state of two concurrent experiments}
\end{figure}

\paragraph{ Re-use of already trained networks}

\noindent In the previous section a general description of experiment use cases has been reported. A specific more detailed information is required by the ``Run'' use case. As known this is the use case selected when a network (for example the MLP model) has been already trained (i.e. after training use case already executed).

\noindent

\begin{figure}  \centering\includegraphics*[width=\textwidth]{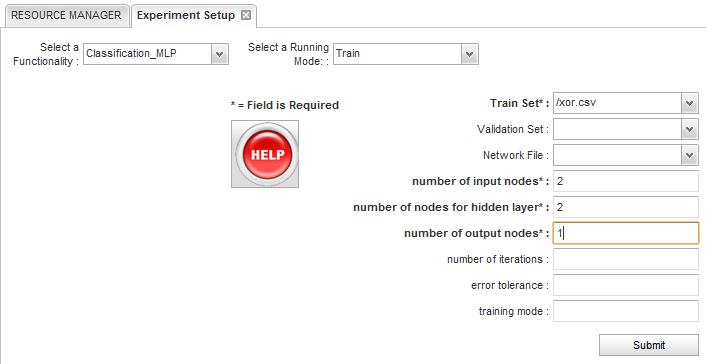}

\caption{ An example of Classification\_MLP training case for the XOR problem}
\end{figure}
\noindent

\begin{figure}  \centering\includegraphics*[width=2.97in]{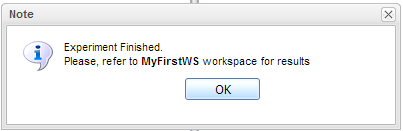}

\caption{ The popup status at the end of the XOR problem experiment}
\end{figure}
\noindent

\begin{figure}  \centering\includegraphics*[width=4.51in]{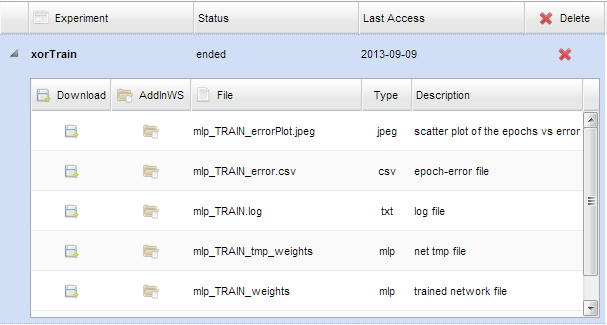}

\caption{ The list of output files after the XOR problem training experiment}
\end{figure}
\noindent

\begin{figure}  \centering\includegraphics*[width=4.17in]{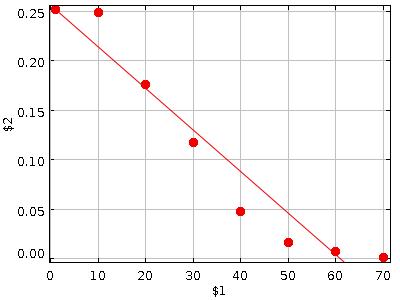}

\caption{ The training error scatter plot mlp\_TRAIN\_errorPlot.jpg downloaded from the experiment output list (x-axis is the training cycle, y-axis is the training mean square error)}
\end{figure}
\noindent The Run case is hence executed to perform scientific experiments on new data. Remember also that the input file does not include ``target'' values. The execution of a Run use case, for its nature, requires special steps in the DAME Suite. These are described in the following.

\noindent

\noindent As first step, we require to have already performed a train case for any experiment, obtaining a list of output files (train or full use cases already executed). In particular in the output list of the train/full experiment there is the file \textit{.mlp}. This file contains the final trained network, in terms of final updated weights of neuron layers, exactly as resulted at the end of the training phase. Depending on the training correctness this file has in practice to be submitted to the network as initial weight file, in order to perform test/run sessions on input data (without target values).

\noindent

\begin{figure}  \centering\includegraphics*[width=3.77in]{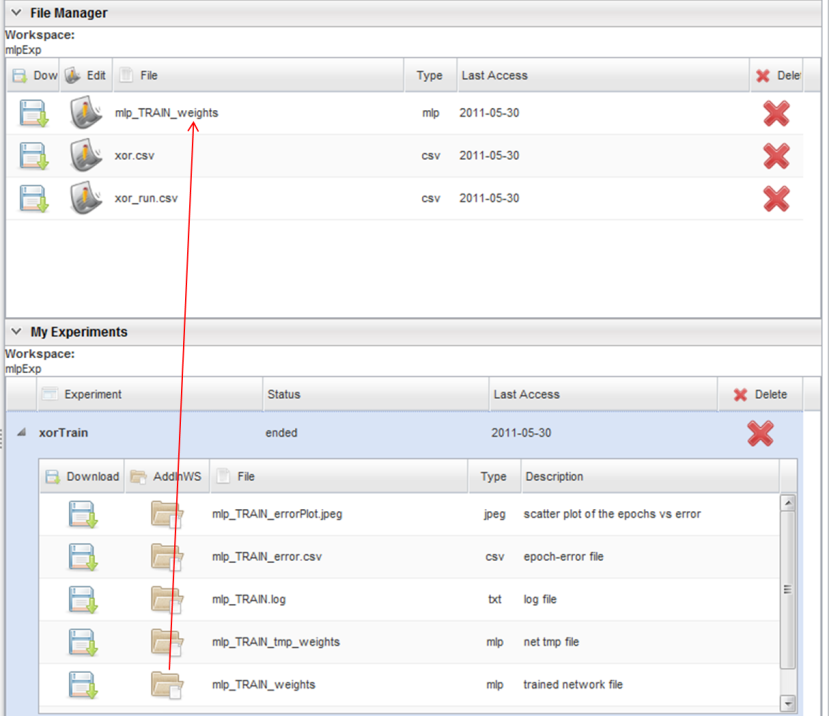}

\caption{ The operation to ``move'' the trained network file in the Workspace input file list}
\end{figure}
\noindent

\noindent To do this, the output weight file must become an input file in the workspace file list, as already explained in section 4.4.4, otherwise it cannot be used as input of Test/Run use case experiment, Fig. 54. Also, the workspace currently active, hosting the experiment we are going to do, must contain a proper input file for Run cases, i.e. without target columns inside.

\noindent So far, the second step is to populate the workspace file list with trained network and Test/Run compliant input files and then to configure and execute the test experiment (see Fig. 55)

\noindent

\begin{figure}  \centering\includegraphics*[width=4.78in]{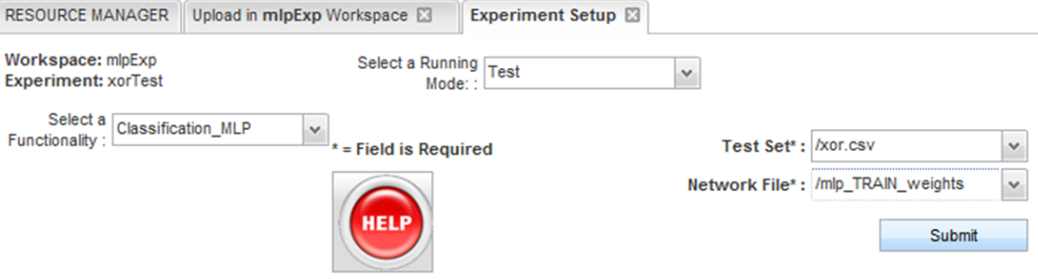}

\caption{the configuration for the Run use case in the XOR problem}
\end{figure}
\noindent

\begin{figure}  \centering\includegraphics*[width=\textwidth]{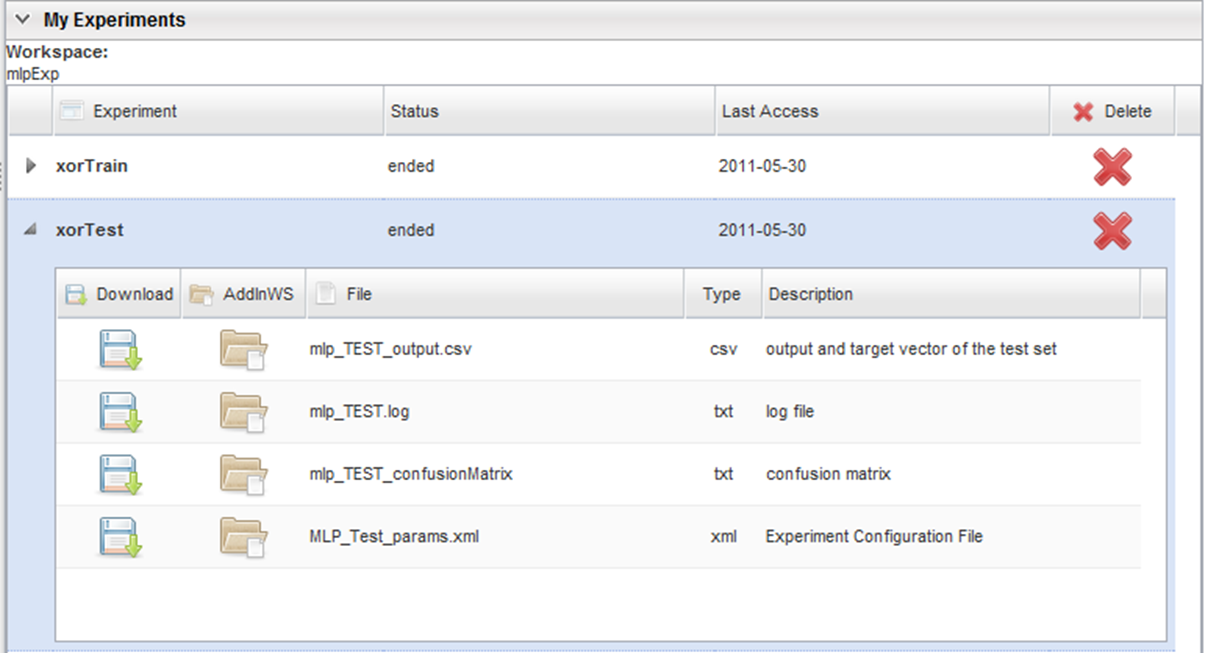}

\caption{ the output of the TEST use case experiment in the XOR problem}
\end{figure}
\noindent

\noindent

\noindent At the end of TEST experiment execution, the experiment output area should contain a list of output files, as shown inFig. 54.

\noindent

\noindent Also the same file \textit{.mlp} should be selected as Network file input in case you want to execute another training (TRAIN/FULL cases) phase, for example when first training session ended in an unsuccessful or insufficient way. In this cases the user can execute more training experiments, starting learning from the previous one, by resuming the trained weight matrix as input network for future training sessions.. This operation is the so-called ``\textit{resume training}'' phase of a neural network.

\noindent

\noindent Of course, the same XOR problem could be also solved by using another functionality - model couple (such as Regression\_FMLPGA).

\noindent We remind the user to consult, when available, the related model specific documentation and manuals, available from the header menu of the webapp, the beta intro web page or the machine learning web page of the official DAME website.

\section{ Abbreviations and Acronyms}
\begin{center}
\begin{tabular}{|p{0.7in}|p{2.5in}|} \hline
\textbf{A \& A} & \textbf{Meaning} \\ \hline
ARFF & Attribute Relation File Format \\ \hline
ASCII & American Standard Code for Information Interchange \\ \hline
BoK & Base of Knowledge \\ \hline
CE & Cross Entropy \\ \hline
CSOM & Clustering Self Organizing Maps \\ \hline
CSV & Comma Separated Values \\ \hline
DAME & DAta Mining \& Exploration  \\ \hline
DM & Data Mining \\ \hline
DMS & Data Mining Suite \\ \hline
FITS & Flexible Image Transport System \\ \hline
GRID & Global Resource Information Database \\ \hline
GSOM & Gated Self Organizing Maps \\ \hline
GUI & Graphical User Interface \\ \hline
INAF & IstitutoNazionale di Astrofisica \\ \hline
JPEG & Joint Photographic Experts Group \\ \hline
MLP & Multi Layer Perceptron \\ \hline
FMLPGA & Fast MLP with Genetic Algorithms \\ \hline
MLPQNA & MLP with Quasi Newton Algorithm \\ \hline
OAC & Osservatorio Astronomico di Capodimonte \\ \hline
PC & Personal Computer \\ \hline
SOFM & Self Organizing Feature Maps \\ \hline
SOM & Self Organizing Maps \\ \hline
URI & Uniform Resource Indicator \\ \hline
VO & Virtual Observatory \\ \hline
XML & eXtensible Markup Language \\ \hline
\end{tabular}
\end{center}

\section*{Acknowledgments}

\noindent \textbf{}

\noindent The DAME program has been funded ~by the Italian Ministry of Foreign Affairs, the European project~\textit{VOTECH}~(Virtual Observatory Technological Infrastructures) and by the Italian~\textit{PON-S.Co.P.E}. Leaders of the project are prof. G. Longo and prof. G.S. Djorgovski.

\noindent

\noindent The current release of the data mining Suite is a miracle due mainly to the incredible effort of (in alphabetical order):

\noindent

\noindent \textit{Giovanni Albano, Stefano Cavuoti, Giovanni d'Angelo, Alessandro Di Guido, Francesco Esposito, Pamela Esposito, Michelangelo Fiore, Mauro Garofalo, Marisa Guglielmo, Omar Laurino,  Francesco Manna, Alfonso Nocella, Sandro Riccardi, Bojan Skordovski, Civita Vellucci}

\noindent

\noindent We want to really thank all actors who contribute and sustain our common efforts to make the whole DAME Program a reality, coming from University Federico II of Naples, INAF Astronomical Observatory of Capodimonte and Californian Institute of Technology.

\vskip 5cm

\noindent \textbf{\includegraphics*[width=\textwidth]{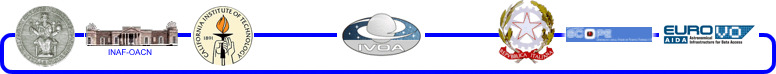}}


\addcontentsline{toc}{section}{References}
\begin{thebibliography}{99}
\bibitem{1} Breiman, L., 2001, Machine Learning, 45 : 5--32. doi:10.1023/A:1010933404324.
\bibitem{2} Brescia, M., Cavuoti, S., \& Longo, G., 2015, MNRAS, 450, 3893.
\bibitem{3} Brescia, M., Cavuoti, S., Longo, G., \& De Stefano, V., 2014, A\&A, 568, A126.
\bibitem{4} Brescia, M., Cavuoti, S., Longo, G., et al. 2014, PASP, 126, 783.
\bibitem{5} Brescia, M., Cavuoti, S., D'Abrusco, R., Longo, G., \& Mercurio, A., 2013, APJ, 772, 140.
\bibitem{6} Brescia, M., Cavuoti, S., Paolillo, M., Longo, G., \& Puzia, T., 2012, MNRAS, 421, 1155.
\bibitem{7} Brescia, M., Longo, G., Djorgovski, G.S., et al. 2010, Astrophysics Source Code Libraryascl:1011.006.
\bibitem{8} Cavuoti, S., 2015, Data-Rich Astronomy: Mining Synoptic Sky Surveys, LAMBERT Academic Publishing, ISBN: 978-3-659-68311-4, 260pp.
\bibitem{9} Cavuoti, S., Brescia, M., Tortora, C., et al., 2015, MNRAS, 452, 3100.
\bibitem{10} Cavuoti, S., Brescia, M., De Stefano, V., \& Longo, G., 2015, Experimental Astronomy, 39, 45.
\bibitem{11} Cavuoti, S., Brescia, M., D'Abrusco, R., Longo, G., \& Paolillo, M., 2014, MNRAS, 437, 968.
\bibitem{12} Cavuoti, S., Brescia, M., Longo, G., \& Mercurio, A., 2012, A\&A, 546, A13.
\bibitem{13} Chang, C.C. and Lin, C.J., 2011, ACM Transactions on Intelligent Systems and Technology, 2:27:1--27:27.
\bibitem{14} de Jong, J.T.A., Verdoes Kleijn, G.A., Boxhoorn, D.R., et al. 2015, A\&A, 582, A62.
\bibitem{15} Deng, D. and Kasabov, N. Neural Networks, 2000. IJCNN 2000, Proceedings of the IEEE-INNS-ENNS International Joint Conference on, Como, 2000, pp. 3-8 vol.6.
\bibitem{16} D'Isanto, A., Cavuoti, S., Brescia, M., et al., 2016, MNRAS, 457, 3119.
\bibitem{17} Esposito, F., 2013, bachelor thesis.
\bibitem{18} Holland, J.H., 1975, Adaptation in Natural and Artificial Systems. University of Michigan Press, Ann Arbor
\bibitem{19} Kelley, C. T., 1999,~Iterative Methods for Optimization, SIAM Frontiers in Applied Mathematics, no 18.
\bibitem{20} Kohonen, T., 1982, Biological Cybernetics,~43: 59--69.
\bibitem{21} MacQueen, J. B., 1967, Proceedings of 5-th Berkeley Symposium on Mathematical Statistics and Probability, Berkeley, University of California Press, 1:281-297.
\bibitem{22} Masters, D., Capak, P., Stern, D., et al., 2015, APJ, 813, 53.
\bibitem{23} Rosenblatt, F., 1961, Principles of Neurodynamics: Perceptrons and the Theory of Brain Mechanisms. Spartan Books, Washington DC.
\bibitem{24} Tortora, C., La Barbera, F., Napolitano, N.R., et al., 2016, MNRAS, 457, 2845.

\end{thebibliography}
\end{document}